\documentclass{aa}  
\usepackage{rotating}
\usepackage{graphicx}
\usepackage{txfonts}
\newcommand{\cm}{cm$^{-1}$}
\newcommand{\boldtext}{}

\titlerunning{Ion--molecule reactions}
\authorrunning{Mladenovi\'c \& Roueff}

\begin{document} 

\title{Ion-molecule reactions involving HCO$^+$ and N$_2$H$^+$:
       Isotopologue equilibria from new theoretical calculations 
       and consequences for interstellar isotope fractionation }
%   \subtitle{I. Overviewing the $\kappa$-mechanism}

   \author{M. Mladenovi\'c
          \inst{1}
          \and
          E. Roueff 
          \inst{2}
%\fnmsep\thanks{Just to show the usage
%          of the elements in the author field}
          }

   \institute{Universit{\' e} Paris-Est,
          Laboratoire Mod{\' e}lisation et Simulation
          Multi Echelle, MSME UMR 8208 CNRS, 5 bd Descartes,
          77454 Marne la Vall{\' e}e, France
\\
              \email{Mirjana.Mladenovic@u-pem.fr}
         \and
             {LERMA and UMR 8112 du CNRS, 
          Observatoire de Paris, Section de Meudon,
          Place J. Janssen, 92195 Meudon, France}
\\
             \email{evelyne.roueff@obspm.fr}
%   \thanks{The university of heaven temporarily does not accept e-mails}
             }

   \date{Received February 28th, 2014; accepted April 14th, 2014}

% \abstract{}{}{}{}{} 
% 5 {} token are mandatory
 
  \abstract
   {}
%    Several fractionation reactions involved in the $^{12}$C/$^{13}$C, 
%    $^{16}$O/$^{18}$O,
%    and $^{14}$N/$^{15}$N balance are investigated in great detail.  }
   {{\boldtext
     We revisit with new augmented accuracy the theoretical dynamics of basic
     isotope exchange reactions involved in the $^{12}$C/$^{13}$C, 
     $^{16}$O/$^{18}$O, and $^{14}$N/$^{15}$N balance
     because these reactions have already been studied experimentally
     in great detail. 
     }
%%    Several fractionation reactions involved in the $^{12}$C/$^{13}$C, 
%%    $^{16}$O/$^{18}$O,
%%    and $^{14}$N/$^{15}$N balance are investigated in great detail 
%%    in order to gain an initial understanding of 
%%    some features governing the dynamics of
%%    the isotope exchange reactions.}
    }
%   Our goal is to address questions related to the dynamics of
%   the isotope exchange reactions in the molecular universe.}
%   {Our general goal is to understand the role of
%    the isotope exchange reactions in the molecular universe.}
   {Electronic structure methods were employed to explore
    potential energy surfaces, full-dimensional rovibrational calculations 
    %to compute numerically exact rovibrational energy levels, 
    to compute rovibrational energy levels that are numerically exact,
    and chemical network models to estimate the abundance ratios under
    interstellar conditions.}
   {New exothermicities, derived for HCO$^+$ reacting with CO, 
    provide rate coefficients
    markedly different from previous theoretical values 
    in particular at low temperatures, resulting in
    new abundance ratios relevant for carbon chemistry networks.
    {\boldtext
    In concrete terms, we obtain 
    a reduction in the abundance of H$^{12}$C$^{18}$O$^+$ 
    and an increase in the abundance of 
    H$^{13}$C$^{16}$O$^+$ and D$^{13}$C$^{16}$O$^+$.
    }
    In all studied cases, the reaction of the ion with 
    a neutral polarizable molecule proceeds
    through the intermediate proton-bound complex found to be very stable.
    {\boldtext
     For the complexes
         {OCH$^+\cdots$CO},   %%{OCHCO$^+$(TS)},  
          {OCH$^+\cdots$OC}, 
          {COHOC$^+$}, 
          {N$_2 \cdots$HCO$^+$}, 
          {N$_2$H$^+\cdots$OC}, and {N$_2$HN$_2^+$}, we also calculated
     vibrational frequencies and dissociation energies.
    }
} 
   {The linear proton-bound complexes possess sizeable dipole moments, which
    may facilitate their detection.}


   \maketitle
%
%________________________________________________________________

\section{Introduction}
\label{sec:intro}

Isotopic fractionation reactions have already been invoked by \cite{watson76a} and \cite{dalgarno76} 
to explain the enrichment of heavy isotopes of molecules in dark cold interstellar cloud environments. 
The exothermicity involved in the isotopic exchange reaction directly depends on the difference 
of the zero-point energies (ZPE) between the two isotopes, if one assumes that the reaction proceeds 
in the ground-rovibrational states of both the reactant and product molecule. 
This assumption has been questioned for the reaction H$_3^+$+HD$\rightleftharpoons$H$_2$D$^+$+H$_2$, 
where some rotational excitation in H$_2$ may reduce 
the efficiency of the reverse reaction \citep{pagani92,hugo09}. 

In this paper we revisit some fractionation reactions involved in the $^{12}$C/$^{13}$C, $^{16}$O/$^{18}$O, 
and $^{14}$N/$^{15}$N balance by reinvestigating the potential energy surfaces involved 
in the isotopic exchange reactions. 
Within the Born-Oppenheimer approximation, a single nuclear-mass-independent potential energy surface 
(PES)
is considered for all isotopic variants of molecules under consideration. 
The nuclear motions are introduced subsequently and isotopologues, 
molecules of different isotopic compositions and thus different masses, 
possess different rotational constants, 
different vibrational frequencies, and
different ground-state (zero-point) vibrational energies,
in other words, different thermodynamic properties \citep{urey47}.
Differences in zero-point energies can become important 
under cool interstellar cloud conditions where
molecules rather undergo isotopic exchange (fractionation) 
than react chemically.
This thermodynamic effect may result in 
isotopologue abundance ratios (significantly) 
% enhanced/reduced with respect to 
deviating from the elemental isotopic ratios.
Knowledge of the abundance ratios may in return provide
valuable information on molecular processes 
at low collision energies.

As far as astrophysical models are concerned, $^{13}$C and $^{18}$O 
isotopic fractionation studies involving CO and HCO$^+$ 
\citep{lebourlot93,liszt07, roellig13, maret13} are based 
on the pionneering paper by \cite{langer84}, 
who referred to the experimental studies by \cite{smith80a} 
and used theoretical spectroscopic parameters for the isotopic variants 
of HCO$^+$ reported by \cite{henning77}. 
\cite{lohr98} derived the harmonic frequencies and equilibrium rotational constants
for CO, HCO$^+$, and HOC$^+$ 
at the configuration interaction (including single and double excitations) 
level of theory (CISD/6-31G**)
and tabulated reduced partition function ratios
and isotope exchange equilibrium constants
for various isotope exchange reactions between CO and HCO$^+$.
Surprisingly, this paper has not 
received much attention in the astrophysical literature, 
and its conclusions have never been applied. 

The studies of \cite{langer84} and \cite{lohr98}
led to qualitatively different conclusions regarding the following fractionation reaction:
\begin{eqnarray}
{^{13}\mathrm{C}}{^{16}\mathrm{O}}  + \mathrm{H}^{12}\mathrm{C}{^{18}\mathrm{O}}^+  
%%& \rightleftharpoons &
\rightarrow
 \mathrm{H}^{13}\mathrm{C}{^{16}\mathrm{O}}^+  
+ {^{12}\mathrm{C}}{^{18}\mathrm{O}}                     
%\nonumber\\
%& &
+ \Delta E
,
~~
\label{reaction_diff} 
\end{eqnarray}
\noindent
which was found to be endothermic with $\Delta E/k_{\mathrm{B}}=-5$ K 
by \cite{langer84}
and exothermic with $\Delta E/k_{\mathrm{B}} = 12.5$ K by \cite{lohr98}, 
where $k_{\mathrm{B}}$ is the Boltzmann constant.
To clear up this discrepancy, 
we carried out numerically exact calculations for
the vibrational ground state of
HCO$^+$ using a potential energy surface previously
developed by \cite{mladenovic98b}.
Our calculations gave $\Delta E/k_{\mathrm{B}} = 11.3$ K
for reaction (\ref{reaction_diff}), in good agreement 
with the harmonic value of \cite{lohr98}.
In addition, we noticed that 
the $\Delta E/k_{\mathrm{B}}$ values of \citet{henning77} 
for the reactions
\begin{eqnarray}
{^{13}\mathrm{C}}{^{16}\mathrm{O}}  
+ \mathrm{H}^{12}\mathrm{C}{^{16}\mathrm{O}}^+  
& \rightleftharpoons &
 \mathrm{H}^{13}\mathrm{C}{^{16}\mathrm{O}}^+  
+ {^{12}\mathrm{C}}{^{16}\mathrm{O}}                     
~~~~
\label{reaction2} 
\end{eqnarray}
\noindent
and 
\begin{eqnarray}
{^{12}\mathrm{C}}{^{18}\mathrm{O}}                     
+ \mathrm{H}{^{12}\mathrm{C}}{^{16}\mathrm{O}}^+  
& \rightleftharpoons &
\mathrm{H}{^{12}\mathrm{C}}{^{18}\mathrm{O}}^+ 
+ {^{12}\mathrm{C}}{^{16}\mathrm{O}}                     
~~~~
\label{reaction3} 
\end{eqnarray}
\noindent
were quoted as 17$\pm$1 K and 7$\pm$1 K by \cite{smith80a} 
and as 9 and 14 K by \cite{langer84}. 
Reconsidering the original values of \cite{henning77}, 
we found that \cite{langer84} permuted 
the zero-point energies 
for H$^{13}$C$^{16}$O$^+$ and H$^{12}$C$^{18}$O$^+$ 
in Table 2 of their paper.
From the original spectroscopic parameters of \cite{henning77}, we derive
$\Delta E/k_{\mathrm{B}} = 10.2$ K for reaction (\ref{reaction_diff}), 
in good agreement
with our result and the result of \cite{lohr98}. 

The permutation of the zero-point vibrational energies of
H$^{13}$C$^{16}$O$^+$ and H$^{12}$C$^{18}$O$^+$ affects
the exothermicities and rate coefficients summarized 
in Table 1 and Table 3 of the paper by \cite{langer84}.
These data are actually incorrect
for all isotope fractionation reactions CO+HCO$^+$, 
except for
\begin{eqnarray}
{^{13}\mathrm{C}}{^{18}\mathrm{O}}  
+ \mathrm{H}^{12}\mathrm{C}{^{16}\mathrm{O}}^+  
& \rightleftharpoons &
%%& \rightarrow&
 \mathrm{H}^{13}\mathrm{C}{^{18}\mathrm{O}}^+  
+ {^{12}\mathrm{C}}{^{16}\mathrm{O}}                     
%%%%+ \Delta E
.
~~~~
\label{reaction_isto} 
\end{eqnarray}
\noindent
The rate coefficients reported by \cite{langer84} are still widely used when including
isotopes such as $^{13}$C and $^{18}$O into chemical (molecular) networks
\citep{maret13,roellig13}.
With these points in mind, our goal is to provide
reliable theoretical estimates for the zero-point vibrational energies
first of H/DCO$^+$ and to derive proper rate coefficients for the related
fractionation reactions.
Our improved results for the exothermicities and rate coefficients
are summarized
in Tables \ref{table_kelvin} and \ref{table_rate}.

\cite{henning77} also reported spectroscopic parameters for
various isotopic variants of N$_2$H$^+$.
This was our initial motivation to expand the present study 
to ion-molecule reactions between N$_2$H$^+$ and N$_2$.
$^{15}$N fractionation in dense interstellar clouds has been first 
considered by \cite{terzieva00}, 
who referred to
the experimental information of the selected ion flow-tube 
(SIFT) studies at low temperatures 
of \cite{adams81}. 

{\boldtext
The reactions discussed in this paper,
CO+HCO$^+$ and N$_2$+HN$_2^+$,
% HCO$^+$ with CO and N$_2$H$^+$ with N$_2$, 
are the most obvious candidates 
for isotopic fractionation. 
In addition, they have been studied in the laborataory, which allows 
a detailed discussion.
A similar reaction has been invoked for CN \citep{milam09}, 
but no experimental and/or theoretical information is available there.
}

In the Langevin model, 
the long-range contribution to the intermolecular potential 
is described by the isotropic interaction between the charge of the ion and 
the induced dipole of the neutral.
Theoretical approaches based on this standard assumption may qualitatively explain 
the behaviour of the association rates. However, they generally provide
rate coefficients that are higher than experimental results \citep{langer84}.
The rate coefficients for ion-molecule reactions are quite constant
at higher temperatures but increase rapidly at lower temperatures.
The latter feature is an indication of barrierless potential energy surfaces.
The electrostatic forces are always attractive and can be experienced
over large distances even at extremely low temperatures relevant for
dark cloud enviroments.
Short-range forces appear in closer encounters
of interacting particles and may (prominently) influence the overall reaction rate.
To explore the short-range effects we also undertake a study of linear
proton-bound ionic complexes arising in the reactions involving
HCO$^+$, HOC$^+$, and N$_2$H$^+$ with CO and N$_2$, which are common
interstellar species.

Our theoretical approach is described in Sect. \ref{sec:calculations}.
The specific  aspects of the fractionation reactions of
HCO$^+$ and HOC$^+$ with CO are reanalysed in Sect. \ref{sec:hco+}
and the fractionation reactions N$_2$H$^+$+N$_2$ in Sect. \ref{sec:nnh+}.    
We discuss the equilibrium constants and rate coefficients of
CO+HCO$^+$/HOC$^+$ in Sect. \ref{sec_discussion:hco+}, providing
the astrochemical implications of the new exothermicities
in Sect. \ref{discussion:astro}.
The isotope fractionation reactions N$_2$H$^+$+N$_2$ are considered
including the nuclear spin angular momentum selection rules
in Sect. \ref{sec_discussion:nnh+}.
The linear proton-bound cluster ions are analysed
in Sect. \ref{discussion:pes}.
Our concluding remarks are given in Sect. \ref{sec:conclusion}.
%________________________________________________________________

\section{Calculations}
\label{sec:calculations}

The global three-dimensional potential energy surfaces (PES) 
developed by \cite{mladenovic98b} 
for the isomerizing system HCO$^+$/HOC$^+$ 
and by \cite{schmatz97}
for the isoelectronic species N$_2$H$^+$ were used 
in the rovibrational calculations.
These two PESs still provide the most comprehensive
theoretical descriptions of the spectroscopic properties for
HCO$^+$, HOC$^+$, and N$_2$H$^+$ and are valid
up to the first dissociation limit.
Potential energy representations recently developed 
by \cite{spirko08} and 
by \cite{huang10} reproduce the experimental fundamental transitions 
within 11[6] and 4[3] {\cm} for N$_2$H$^+$[N$_2$D$^+$], respectively,
whereas the PES of \cite{schmatz97} predicts  the fundamental transitions
for both N$_2$H$^+$ and N$_2$D$^+$ within 2 {\cm}.

The rovibrational energy levels of HCO$^+$/HOC$^+$ and N$_2$H$^+$
are calculated by a numerically exact quantum
mechanical method, involving no dynamical approximation and
applicable to any potential energy representation.
The computational strategy is based on the discrete variable representation 
of the angular coordinate 
in combination with a sequential diagonalization/truncation
procedure \citep{mladenovic90,mladenovic98b}. 
For both molecular systems,
the rovibrational states are calculated for the total angular momentum
$J=0-15$. % in both parities.
These rovibrational energies are used to evaluate theoretical partition
functions and to model rate coefficients for proton transfer reactions
involving HCO$^+$ and N$_2$H$^+$.

To gain a first insight into dynamical features of ion-mo\-le\-cu\-le reactions,
additional electronic structure calculations were carried out for 
linear proton-bound cluster ions of
HCO$^+$, HOC$^+$, and N$_2$H$^+$ with CO and N$_2$.
The PESs were scanned by means of the coupled cluster method 
with single and double excitations including perturbative corrections 
for triple excitations [CCSD(T)] in combination with the augmented
correlation consistent triple $\zeta$ basis set (aug-cc-pVTZ).
Only valence electrons were correlated.
The ab initio calculations were carried out with the MOLPRO 
\citep{MOLPRO_brief} and CFOUR \citep{CFOUR_brief}
quantum chemistry program packages.

%------------------------------------------------------------------------------------
\section{Results}
\label{sec:results}

The PES of \cite{mladenovic98b} provides a common potential energy representation 
for the formyl cation, HCO$^+$, and the isoformyl cation, HOC$^+$, where 
the local HOC$^+$ minimum is 13\,878 cm$^{-1}$ (166 kJ mol$^{-1}$)
%%(1.7 eV)
above the global HCO$^+$ minimum.
Inclusion of the zero-point energy reduces this separation 
by 640--650 {\cm} for the hydrogen-containing 
isotopologues 
and by 570--580 {\cm} for the deuterium variants.
The angular motion is described by a double-minimum anharmonic potential 
with a non-linear saddle point at 26\,838 {\cm} (321 kJ mol$^{-1}$) 
above the HCO$^+$ minimum, 
such that low-lying states of HCO$^+$ and HOC$^+$ are well separated.

The potential energy surface of \cite{schmatz97} 
for N$_2$H$^+$ (dyazenilium) has two equivalent colinear 
minima as a consequence of the $S_2$ permutation symmetry,
separated by an isomerization barrier 
17\,137 {\cm} (205 kJ mol$^{-1}$)
above the energy of the linear geometries.
Low-lying states of N$_2$H$^+$ are, thus,
localized in one of the two wells.
The double-well symmetry and nuclear spin symmetries
are lifted for mixed nitrogen isotope forms.

%################################################################
\subsection{Reaction of CO with HCO$^+$ and HOC$^+$}
\label{sec:hco+}

The ground-state vibrational energies calculated in this work 
for isotopic variants of HCO$^+$ and HOC$^+$ are collected 
in Table \ref{table_zpe}. 
There we additionally show the harmonic zero-point energy estimates
of \cite{lohr98} 
{\boldtext and 
the anharmonic values of \cite{martin93a} 
available only for three isotopologues,
as well as}
the values obtained by \cite{langer84} 
and in the present work
from the spectroscopic [CI(corr)] parameters of \cite{henning77}.
Our values for CO are computed at the theoretical level
used to construct the potential energy surface for HCO$^+$/HOC$^+$
[CCSD(T)/cc-pVQZ]. 

%----------------------------------------------
%\input{table_zpe}
%---------------------- table ----------------------------------
\begin{table*}
%\tabletypesize{\scriptsize}
\caption{\label{table_zpe}
Zero-point vibrational energies (in {\cm}) of 
isotopologues of CO, HCO$^+$, and HOC$^+$.  }
\vspace{0.025cm}
\centering
%\tablewidth{0pt}
%\tablehead{
%\colhead{Species} & \colhead{This work} & \colhead{{L98}$^a$} & \colhead{{MTL93}$^b$} 
%& \colhead{{LGFA84}$^c$} & \colhead{{HKD77}$^d$} & \colhead{{HH79}$^e$} 
%}
%\startdata
\begin{tabular}{lcccccc}
\hline\hline
 \noalign{\smallskip}
  {Species} & {This work} & {{L98}$^a$} & {{MTL93}$^b$} 
& {{LGFA84}$^c$} & {{HKD77}$^d$} & {{HH79}$^e$} \\
 \noalign{\smallskip}
\hline
 \noalign{\smallskip}
$^{12}$C$^{16}$O  &  1\,079.11 & 1\,131.9 & & 1\,084.8  &       & 1\,081.6\\
$^{13}$C$^{16}$O  &  1\,055.12 & 1\,106.8 & & 1\,060.6  &       & 1\,057.5\\
$^{12}$C$^{18}$O  &  1\,053.11 & 1\,104.8 & & 1\,059.0  &       & 1\,055.5\\
$^{13}$C$^{18}$O  &  1\,028.52 & 1\,078.7 & & 1\,034.0  &       & 1\,030.7\\
 \noalign{\smallskip}
H$^{12}$C$^{16}$O$^+$  & 3\,524.60 & 3\,713.6 & 3\,512.3 & 3\,487.6 & 3\,487.6 \\
H$^{13}$C$^{16}$O$^+$  & 3\,488.24 & 3\,674.6 & 3\,475.89$^f$& 3\,457.0 & 3\,452.0 \\
H$^{12}$C$^{18}$O$^+$  & 3\,494.15 & 3\,681.3 &        & 3\,452.1 & 3\,457.1 \\
H$^{13}$C$^{18}$O$^+$  & 3\,457.16 & 3\,641.6 &        & 3\,421.5 & 3\,421.1 \\
 \noalign{\smallskip}
D$^{12}$C$^{16}$O$^+$  & 2\,944.22 & 3\,096.4 &        &        & 2\,918.9  \\
D$^{13}$C$^{16}$O$^+$  & 2\,905.16 & 3\,055.0 &        &        & 2\,880.4  \\
D$^{12}$C$^{18}$O$^+$  & 2\,912.86 & 3\,063.4 &        &        & 2\,887.7  \\
D$^{13}$C$^{18}$O$^+$  & 2\,873.18 & 3\,021.7 &        &        & 2\,848.5  \\
 \noalign{\smallskip}
H$^{16}$O$^{12}$C$^+$  & 2\,871.08 & 2\,874.3 & 2\,907.4 &        & 2\,934.4    \\
H$^{16}$O$^{13}$C$^+$  & 2\,848.66 & 2\,851.0 &        &        & 2\,911.3    \\
H$^{18}$O$^{12}$C$^+$  & 2\,841.37 & 2\,844.4 &        &        & 2\,905.0    \\
H$^{18}$O$^{13}$C$^+$  & 2\,818.42 & 2\,820.5 &        &        & 2\,881.5    \\
 \noalign{\smallskip}
D$^{16}$O$^{12}$C$^+$  & 2\,357.61 & 2\,365.6 &        &        & 2\,411.6    \\
D$^{16}$O$^{13}$C$^+$  & 2\,334.87 & 2\,341.9 &        &        & 2\,389.0    \\
D$^{18}$O$^{12}$C$^+$  & 2\,326.06 & 2\,335.0 &        &        & 2\,381.1    \\
D$^{18}$O$^{13}$C$^+$  & 2\,302.82 & 2\,311.0 &        &        & 2\,357.8    \\
 \noalign{\smallskip}
\hline
%--------------------------------------------------------------
\end{tabular}
\tablefoot{
\tablefoottext{a}{\cite{lohr98}}
\tablefoottext{b}{\cite{martin93a}}
\tablefoottext{c}{\cite{langer84}}
\tablefoottext{d}{computed from the original data of \cite{henning77}}
\tablefoottext{e}{\cite{huber79}}
\tablefoottext{f}{T.~J. Lee, private communication}
}
\end{table*}
%---------------------------------------------------------------

%----------------------------------------------

%----------------------------------------------
%\input{table_kelvin}
%---------------------- table Dis ----------------------------------
\begin{table*}
%\tabletypesize{\scriptsize}
%\rotate
\caption{\label{table_kelvin}
Zero-point energy differences (in K) between the reactants and products
for the isotope fractionation reactions 
of H/DCO$^+$ and H/DOC$^+$ with CO.  }
\vspace{0.025cm}
\centering
%--------------------------------------------------------------
\begin{tabular}{llrrrrrr}
\hline\hline
\noalign{\smallskip}
 & & \multicolumn{4}{c}{Theory}  & & Exp 
\\
\noalign{\smallskip}
\cline{3-6} \cline{8-8}
\noalign{\smallskip}
{Label}& {Reaction} & {This work$^a$} & {L98$^b$} 
& {HKD77$^c$} & {LGFA84$^d$} &~~& {SA80$^e$} 
\\
\noalign{\smallskip}
\hline
\noalign{\smallskip}
& $^{13}$C$^+$+$^{12}$C$^{16}$O $\rightarrow$ $^{12}$C$^+$+$^{13}$C$^{16}$O 
                      &  34.5 [34.7] & 36.0 & & 35  &~& 40$\pm$6 \\
& $^{13}$C$^+$+$^{12}$C$^{18}$O $\rightarrow$ $^{12}$C$^+$+$^{13}$C$^{18}$O 
                      &  35.4 [35.6] & 37.5 & & 36 & & \\
 \noalign{\smallskip}
\hline
 \noalign{\smallskip}
F1 &
H$^{12}$C$^{16}$O$^+$+$^{13}$C$^{16}$O $ \rightarrow$ H$^{13}$C$^{16}$O$^+$+$^{12}$C$^{16}$O
                                  & 17.8 [17.6] & 20.0 &    16.5 & 9 & & 12$\pm$5\\
F2 &
H$^{12}$C$^{18}$O$^+$+$^{13}$C$^{18}$O $ \rightarrow$ H$^{13}$C$^{18}$O$^+$+$^{12}$C$^{18}$O
                                  &  17.8 [17.6] & 19.5 & 16.4 & 8 &   \\
F3 &
H$^{12}$C$^{16}$O$^+$+$^{12}$C$^{18}$O $ \rightarrow$ H$^{12}$C$^{18}$O$^+$+$^{12}$C$^{16}$O
                                  & 6.4 [6.2]   &  7.5 & 6.2 & 14 & & 15$\pm$5\\
F4 &
H$^{13}$C$^{16}$O$^+$+$^{13}$C$^{18}$O $ \rightarrow$ H$^{13}$C$^{18}$O$^+$+$^{13}$C$^{16}$O
                                  & 6.4 [6.2]   &  7.0 &  6.1 & 13 &  \\
F5 & 
H$^{12}$C$^{16}$O$^+$+$^{13}$C$^{18}$O $ \rightarrow$ H$^{13}$C$^{18}$O$^+$+$^{12}$C$^{16}$O
                                  &  24.2 [23.9] & 27.0 &  22.6 & 22 & \\
F6 &
H$^{12}$C$^{18}$O$^+$+$^{13}$C$^{16}$O $\rightarrow$ H$^{13}$C$^{16}$O$^+$+$^{12}$C$^{18}$O
                                  &  11.4 [11.4] &  12.5& 10.3 & -5 & &  $\le$5  \\
%H$^{13}$C$^{16}$O$^+$+$^{12}$C$^{18}$O$\rightarrow$ H$^{12}$C$^{18}$O$^+$+$^{13}$C$^{16}$O
%                                        & -11.3 & -12.5& 5& & $\le$5  \\
 \noalign{\smallskip}
F1(D)
& D$^{12}$C$^{16}$O$^+$+$^{13}$C$^{16}$O $ \rightarrow$ D$^{13}$C$^{16}$O$^+$+$^{12}$C$^{16}$O
                                  &  21.7 [21.5] & 23.5 &     20.7 &  \\
F2(D)
& D$^{12}$C$^{18}$O$^+$+$^{13}$C$^{18}$O $ \rightarrow$ D$^{13}$C$^{18}$O$^+$+$^{12}$C$^{18}$O
                                  &  21.7 [21.5] & 22.5 &     20.9 &   \\
F3(D)
& D$^{12}$C$^{16}$O$^+$+$^{12}$C$^{18}$O $ \rightarrow$ D$^{12}$C$^{18}$O$^+$+$^{12}$C$^{16}$O
                                  &  7.7 [7.5]  &  8.5 &     7.3 &   \\
F4(D)
& D$^{13}$C$^{16}$O$^+$+$^{13}$C$^{18}$O $ \rightarrow$ D$^{13}$C$^{18}$O$^+$+$^{13}$C$^{16}$O
                                  &  7.7 [7.5]  &  7.5 &     7.5 &   \\
F5(D)
& D$^{12}$C$^{16}$O$^+$+$^{13}$C$^{18}$O $ \rightarrow$ D$^{13}$C$^{18}$O$^+$+$^{12}$C$^{16}$O
                                  &  29.4 [29.0] &  31.0&     28.2 &   \\
%D$^{13}$C$^{16}$O$^+$+$^{12}$C$^{18}$O$\rightarrow$ D$^{12}$C$^{18}$O$^+$+$^{13}$C$^{16}$O
%                                        &  -13.9&  -14.5    &  &   \\
F6(D)
& D$^{12}$C$^{18}$O$^+$+$^{13}$C$^{16}$O $\rightarrow$ D$^{13}$C$^{16}$O$^+$+$^{12}$C$^{18}$O
                                  &   14.0 [14.0]&   15.0    &   13.4&  \\
 \noalign{\smallskip}
I1 & H$^{16}$O$^{12}$C$^+$+$^{13}$C$^{16}$O $\rightarrow$ H$^{16}$O$^{13}$C$^+$+$^{12}$C$^{16}$O
                                        &  -2.3 [-2.4] & -2.5      &   -1.5 & \\
I2 & H$^{18}$O$^{12}$C$^+$+$^{13}$C$^{18}$O $\rightarrow$ H$^{18}$O$^{13}$C$^+$+$^{12}$C$^{18}$O
                                        & -2.4 [-2.6]  & -3.0      &   -1.7 &    \\
I3 & H$^{16}$O$^{12}$C$^+$+$^{12}$C$^{18}$O $ \rightarrow$ H$^{18}$O$^{12}$C$^+$+$^{12}$C$^{16}$O
                                        &  5.3 [5.2]  & 4.0       &   4.7 &    \\
I4 & H$^{16}$O$^{13}$C$^+$+$^{13}$C$^{18}$O $ \rightarrow$ H$^{18}$O$^{13}$C$^+$+$^{13}$C$^{16}$O
                                        &   5.2 [5.0] & 3.5     &   4.5 &  \\
I5 & H$^{16}$O$^{12}$C$^+$+$^{13}$C$^{18}$O $ \rightarrow$ H$^{18}$O$^{13}$C$^+$+$^{12}$C$^{16}$O
                                        &   3.0 [2.6] & 1.0 &  3.0 &    \\
I6 & H$^{18}$O$^{12}$C$^+$+$^{13}$C$^{16}$O $ \rightarrow$ H$^{16}$O$^{13}$C$^+$+$^{12}$C$^{18}$O
                                        &  -7.6 [-7.6] & -6.5  & -6.2 &    \\
%H$^{16}$O$^{13}$C$^+$+$^{12}$C$^{18}$O$ \rightarrow$ H$^{18}$O$^{12}$C$^+$+$^{13}$C$^{16}$O
%                                        &   7.6 & 6.5 & &    \\
 \noalign{\smallskip}
I1(D)
& D$^{16}$O$^{12}$C$^+$+$^{13}$C$^{16}$O $\rightarrow$ D$^{16}$O$^{13}$C$^+$+$^{12}$C$^{16}$O
                                        & -1.8 [-2.0] & -2.0 &  -2.1 & \\
I2(D)
& D$^{18}$O$^{12}$C$^+$+$^{13}$C$^{18}$O $\rightarrow$ D$^{18}$O$^{13}$C$^+$+$^{12}$C$^{18}$O
                                        &  -1.9 [-2.1]& -3.0 &  -2.0 &   \\
I3(D)
& D$^{16}$O$^{12}$C$^+$+$^{12}$C$^{18}$O $ \rightarrow$ D$^{18}$O$^{12}$C$^+$+$^{12}$C$^{16}$O
                                        &   8.0 [7.8]&  5.0 &   6.2 &   \\
I4(D)
& D$^{16}$O$^{13}$C$^+$+$^{13}$C$^{18}$O $ \rightarrow$ D$^{18}$O$^{13}$C$^+$+$^{13}$C$^{16}$O
                                        &   7.8 [7.6]& 4.0 &    6.3 & \\
I5(D)
& D$^{16}$O$^{12}$C$^+$+$^{13}$C$^{18}$O $ \rightarrow$ D$^{18}$O$^{13}$C$^+$+$^{12}$C$^{16}$O
                                        &   6.0 [5.7]& 2.0 &    4.2 &  \\
I6(D)
& D$^{18}$O$^{12}$C$^+$+$^{13}$C$^{16}$O $ \rightarrow$ D$^{16}$O$^{13}$C$^+$+$^{12}$C$^{18}$O
                                        &  -9.8 [-9.8] & -7.0  & -8.4 &    \\
\noalign{\smallskip}
\hline
%--------------------------------------------------------------
\end{tabular}
\tablefoot{
\tablefoottext{a}{
Zero-point energy differences obtained using 
experimental CO zero-point values from \cite{huber79} 
are given in brackets.}
\tablefoottext{b}{\cite{lohr98}}
\tablefoottext{c}{computed from the original data of \cite{henning77}}
\tablefoottext{d}{\cite{langer84}}
\tablefoottext{e}{\cite{smith80a}}
}
\end{table*}
%-----------------------------------------------------------------------

%----------------------------------------------

The isotopologues in Table \ref{table_zpe} are arranged in order of
increasing total molecular mass.
For CO and H/DOC$^+$,
the zero-point energies decrease as the total molecular mass increases,
which is not the case for H/DCO$^+$.
Inspection of the table shows that
the substitution of the central atom by its heavier isotope 
($^{12}$C$\rightarrow$$^{13}$C in H/DCO$^+$
and $^{16}$O$\rightarrow$$^{18}$O in H/DOC$^+$) results in  
a more pronounced decrease
%%%%larger lowering 
of the zero-point energy than 
the isotopic substitution of the terminal atom
($^{16}$O$\rightarrow$$^{18}$O in H/DCO$^+$
and $^{12}$C$\rightarrow$$^{13}$C in H/DOC$^+$).
This feature shared by H/DCO$^+$ and  H/DOC$^+$ in Table \ref{table_zpe}
is easy to rationalize
since a central atom substitution affects all three vibrational frequencies.

The zero-point energy differences for the proton transfer reactions
CO+HCO$^+$/HOC$^+$  are listed in Table \ref{table_kelvin}.
The reactions involving the formyl cation are labelled with F
and the reactions involving the isoformyl cation with I.
The deuterium variant of reaction F1 is denoted by F1(D)
and similar for all other reactions.
The reactions F1, F2, F3, F4, F5, and F6
are numbered as 1\,004, 3\,408, 3\,407, 3\,457, 
3\,406, and 3\,458 by \cite{langer84}.

In Table \ref{table_kelvin}, 
our results, the values rederived from the spectroscopic parameters of Henning et al.   
(column HKD77), and the harmonic values of Lohr (column L98) 
all agree within less than 5 K.
These three data sets 
predict the same direction for all listed reactions, 
whereas Langer et al. (column LGFA84) reported 
reaction F6 as endothermic.
The replacement of our theoretical values for CO 
by the experimental values taken from \cite{huber79} affects the zero-point 
energy differences by at most 0.4\,K.

The general trend seen in Table \ref{table_kelvin} is that 
$^{13}$C is preferentially placed in H/DCO$^+$
and $^{18}$O in H/DOC$^+$.
This is in accordance with Table \ref{table_zpe}, showing
a stronger decrease of the zero-point energy upon isotopic
substitution of the central atom.
The substitution of the two $^{16}$O by $^{18}$O
or the two $^{12}$C by $^{13}$C 
has nearly no influence on the exothermicities,
as seen by comparing $\Delta E$ for reactions F1, F1(D), I1, I1(D)
with $\Delta E$ for reactions F2, F2(D), I2, I2(D)
and silimar for reactions
F3, F3(D), I3, I3(D) versus F4, F4(D), I4, I4(D).
Slightly higher exothermicities 
appear for reactions involving deuterium.
The exothermicities for the reactions with the isoformyl isomers are lower than
for the reactions with the formyl forms.

From the measured forward reaction $k_f$ and backward reaction $k_r$ rate coefficients,
\cite{smith80a} calculated the experimental zero-point energy differences
using
\begin{eqnarray}
%\frac{k_f}{k_r} = K = \exp\left( \frac{\Delta E}{k_{\mathrm{B}} T} \right)
\frac{k_f}{k_r} = K_e = e^{{\Delta E}/{k_{\mathrm{B}} T}}
,
\label{ratio_sa}
\end{eqnarray}
\noindent
where $K_e$ is the equilibrium constant.
The total estimated error on $k_f$ and $k_r$ is 
reported to be $\pm$25\% at 80\,K.
Table \ref{table_kelvin} indicates that
the new/improved theoretical values, and  
the experimental finding for reaction F1 agree 
within the experimental uncertainty.
For reactions F3 and F6, we see that
the theoretical results consistently predict a higher $\Delta E$ value
for H$^{12}$C$^{18}$O$^+$ reacting with $^{13}$C$^{16}$O (reaction F6)
than for H$^{12}$C$^{16}$O$^+$ reacting with $^{12}$C$^{18}$O (reaction F3),
whereas the opposite was derived experimentally. 
Note that \cite{smith80a} reported for 
${^{13}\mathrm{C}}{^{16}\mathrm{O}}$ reacting with                     
$\mathrm{H}{^{12}\mathrm{C}}{^{18}\mathrm{O}}^+$
in addition to reaction F6 also a yield of 10\%  for 
the rearrangement channel 
\begin{eqnarray}
{^{13}\mathrm{C}}{^{16}\mathrm{O}}  
+ \mathrm{H}^{12}\mathrm{C}{^{18}\mathrm{O}}^+  
%% & \rightleftharpoons &
& \rightarrow&
 \mathrm{H}^{13}\mathrm{C}{^{18}\mathrm{O}}^+  
+ {^{12}\mathrm{C}}{^{16}\mathrm{O}}                     
.
\label{reaction_diff_jos} 
\end{eqnarray}
\noindent
%%The completion of the latter transformation is not of a simple proton-transfer type,
The latter transformation is not of a simple proton-transfer type
(but bond-rearrangement type) and
must involve a more complicated chemical mechanism 
probably including an activation energy barrier.

%---------------------------------------------------------
\subsection{Reaction of N$_2$ with N$_2$H$^+$}
\label{sec:nnh+}

%----------------------------------------------
%\input{table_zpe_n2h}
%---------------------- table ----------------------------------
\begin{table*}
\caption{\label{table_zpe_n2h}
Zero-point vibrational energies (in {\cm}) of 
isotopologues of N$_2$ and N$_2$H$^+$.  }
\vspace{0.025cm}
\centering
\begin{tabular}{lcccc}
%\tabletypesize{\scriptsize}
%\rotate
%--------------------------------------------------------------
%\tablewidth{0pt}
%\tablehead{
\hline\hline
 \noalign{\smallskip}
{Species} & {This work} & {{HVL10}$^a$} & {{HKD77}$^b$} 
%%& {{TH00}$^c$} 
& {{HH79}$^c$} \\
 \noalign{\smallskip}
\hline
 \noalign{\smallskip}
$^{14}$N$^{14}$N      &         &       &       &      1\,175.7  \\ 
$^{14}$N$^{15}$N      &         &       &       &      1\,156.0  \\
$^{15}$N$^{15}$N      &         &       &       &      1\,136.0  \\
 \noalign{\smallskip}
H$^{14}$N$^{14}$N$^+$ & 3\,507.79 & 3\,508.6 & 3\,468.6 & \\ %%3\,351& \\
H$^{15}$N$^{14}$N$^+$ & 3\,480.90 & 3\,481.8 & 3\,442.6 & \\ %%3\,326& \\
H$^{14}$N$^{15}$N$^+$ & 3\,486.67 & 3\,487.5 & 3\,447.6 & \\ %%3\,332& \\
H$^{15}$N$^{15}$N$^+$ & 3\,459.44 & 3\,460.4 & 3\,420.9 &      \\
 \noalign{\smallskip}
D$^{14}$N$^{14}$N$^+$ & 2\,921.18 & 2\,917.1 & 2\,892.5 &      \\
D$^{15}$N$^{14}$N$^+$ & 2\,892.33 & 2\,888.3 & 2\,864.7 &      \\
D$^{14}$N$^{15}$N$^+$ & 2\,899.54 & 2\,895.6 & 2\,871.2 &      \\
D$^{15}$N$^{15}$N$^+$ & 2\,870.36 & 2\,866.6 & 2\,843.1 &      \\
 \noalign{\smallskip}
\hline
%--------------------------------------------------------------
\end{tabular}
\tablefoot{
\tablefoottext{a}{computed from the original data of \cite{huang10}}
\tablefoottext{b}{computed from the original data of \cite{henning77}}
%%\tablefoottext{c}{\cite{terzieva00}}
\tablefoottext{c}{\cite{huber79}}
}
\end{table*}
%---------------------------------------------------------------

%----------------------------------------------

The zero-point vibrational energies calculated for N$_2$H$^+$ 
are summarized in Table \ref{table_zpe_n2h}.
In addition to the results obtained for
the potential energy surface of \cite{schmatz97},
Table \ref{table_zpe_n2h} also provides the values we derived from the spectroscopic
parameters of \cite{huang10} (column HVL10) and 
of \cite{henning77} (column HKD77). 
%as well as
%the harmonic estimates from the scaled M{\o}ller-Plesset calculations 
%(column TH00) by \cite{terzieva00}.
The values for N$_2$  are taken from \cite{huber79}.
The zero-point energy differences are given in Table \ref{table_kelvin_n2h}.
As seen there, our results agree with the values obtained from 
the spectroscopic parameters of \cite{huang10}
within 0.4 K.
The reactions involving diazenylium (or dinitrogen monohydride cation)
are labelled with D in Table \ref{table_kelvin_n2h}.

%----------------------------------------------
%\input{table_kelvin_n2h}
%--------------------------------------------------------------
\begin{table*}
%\tabletypesize{\scriptsize}
%\rotate
\caption{\label{table_kelvin_n2h}
Zero-point energy differences (in K) between the reactants and products
for the isotope fractionation reactions 
of N$_2$H$^+$ with N$_2$.  }
\vspace{0.025cm}
\centering
\begin{tabular}{llrrrrrr}
\hline\hline
\noalign{\smallskip}
 & & \multicolumn{4}{c}{Theory}  & & Exp %\multicolumn{1}{c}{Exp}
\\
\noalign{\smallskip}
\cline{3-6} \cline{8-8}
\noalign{\smallskip}
{Label} & {Reaction} & {This work} 
& {HVL10$^a$}
& {TH00$^b$}
& {HKD77$^c$}
&~~
& {AS81$^d$}
\\
%}
%\startdata
%-------------------------------------------------------------------
 \noalign{\smallskip}
\hline
 \noalign{\smallskip}
D1 &
$^{14}$N$_2$H$^+$+$^{15}$N$_2$ $\rightarrow$ $^{15}$N$_2$H$^+$+$^{14}$N$_2$
                                                & 12.4  & 12.0 &      & 11.3   & & 13$\pm$3 \\
D2 & 
$^{14}$N$_2$H$^+$+$^{15}$N$^{14}$N $\rightarrow$ $^{14}$N$^{15}$NH$^+$+$^{14}$N$_2$
                                                & 10.3  & 10.1 & 10.7 & 9.0 && 9$\pm$3 \\
D3 & 
$^{14}$N$_2$H$^+$+$^{15}$N$^{14}$N $\rightarrow$ $^{15}$N$^{14}$NH$^+$+$^{14}$N$_2$
                                                & 2.0   &  1.9& 2.25   & 1.9 &      \\
D4 & 
$^{14}$N$^{15}$NH$^+$+$^{15}$N$_2$ $\rightarrow$ $^{15}$N$_2$H$^+$+$^{15}$N$^{14}$N
                                                & 2.0   &  1.9 &       & 2.3 && 9$\pm$3 \\
D5 & 
$^{15}$N$^{14}$NH$^+$+$^{15}$N$_2$ $\rightarrow$ $^{15}$N$_2$H$^+$+$^{15}$N$^{14}$N
                                                & 10.3  &  10.1 &       & 9.4  \\
D6 & 
$^{15}$N$^{14}$NH$^+$+$^{15}$N$^{14}$N $\rightarrow$ $^{14}$N$^{15}$NH$^+$+$^{14}$N$^{15}$N
                                                &  8.3  &  8.2 &       & 7.1  \\
 \noalign{\smallskip}
D1(D)
& $^{14}$N$_2$D$^+$+$^{15}$N$_2$ $\rightarrow$ $^{15}$N$_2$D$^+$+$^{14}$N$_2$
                                                & 15.9  & 15.6&         & 14.0 \\
D2(D)
& $^{14}$N$_2$D$^+$+$^{15}$N$^{14}$N $\rightarrow$ $^{14}$N$^{15}$ND$^+$+$^{14}$N$_2$
                                                &  13.2 & 13.1 &        & 11.7\\
D3(D)
& $^{14}$N$_2$D$^+$+$^{15}$N$^{14}$N $\rightarrow$ $^{15}$N$^{14}$ND$^+$+$^{14}$N$_2$
                                                &  2.8  &  2.6 &       & 2.4 \\
D4(D)
& $^{14}$N$^{15}$ND$^+$+$^{15}$N$_2$ $\rightarrow$ $^{15}$N$_2$D$^+$+$^{15}$N$^{14}$N
                                                &  2.8  &  2.5 &       & 2.3  \\
D5(D)
& $^{15}$N$^{14}$ND$^+$+$^{15}$N$_2$ $\rightarrow$ $^{15}$N$_2$D$^+$+$^{15}$N$^{14}$N
                                                & 13.1  & 12.9 &        & 11.6 \\
D6(D)
& $^{15}$N$^{14}$ND$^+$+$^{15}$N$^{14}$N $\rightarrow$ $^{14}$N$^{15}$ND$^+$+$^{14}$N$^{15}$N
                                                & 10.4  &  10.4&        & 9.2  \\
 \noalign{\smallskip}
\hline
%--------------------------------------------------------------
\end{tabular}
\tablefoot{
\tablefoottext{a}{computed from the original data of \cite{huang10}}
\tablefoottext{b}{\cite{terzieva00}}
\tablefoottext{c}{computed from the original data of \cite{henning77}}
\tablefoottext{d}{\cite{adams81}}
}
\end{table*}
%-----------------------------------------------------------------------

%----------------------------------------------

The $^{14}$N/$^{15}$N substitution at the central-atom position
lowers the zero-point energy more than the terminal-atom substitution
(Table \ref{table_zpe_n2h}), such that
$^{15}$N preferentially assumes the central position in N-N-H$^+$ in all
reactions of N$_2$H$^+$ with N$_2$ in Table \ref{table_kelvin_n2h}.
The exothermicities are found to be slightly higher 
for the reactions involving deuterium. 

In Table \ref{table_kelvin_n2h}, the experimental (SIFT) results
of \cite{adams81} are listed as given in their paper.
Note, however, that the elementary isotope fractionation reactions
D2 and D3
\begin{eqnarray}
{^{14}\mathrm{N}}_2{\mathrm{H}}^+
+ {^{15}\mathrm{N}}{^{14}\mathrm{N}}
\mathrel{\mathop{\rightleftharpoons}^{k_2}_{k_{-2}}} 
% \mathrel{\mathop{\nearrow}^{k_2}_{k_{-2}}} 
{^{14}\mathrm{N}}{^{15}\mathrm{N}}{\mathrm{H}}^+
+ {^{14}\mathrm{N}}_2,
\label{reaction_n2h_2} \\
{^{14}\mathrm{N}}_2{\mathrm{H}}^+
+ {^{15}\mathrm{N}}{^{14}\mathrm{N}}
 \mathrel{\mathop{\rightleftharpoons}^{k_3}_{k_{-3}}} 
{^{15}\mathrm{N}}{^{14}\mathrm{N}}{\mathrm{H}}^+
+ {^{14}\mathrm{N}}_2,
\label{reaction_n2h_3}
\end{eqnarray}
\noindent
involve common reactants, whereas reactions D4 and D5
\begin{eqnarray}
{^{14}\mathrm{N}}{^{15}\mathrm{N}}{\mathrm{H}}^+
+ {^{15}\mathrm{N}}_2
 \mathrel{\mathop{\rightleftharpoons}^{k_4}_{k_{-4}}} 
{^{15}\mathrm{N}}_2{\mathrm{H}}^+
+ {^{15}\mathrm{N}}{^{14}\mathrm{N}},
\label{reaction_n2h_4} \\
{^{15}\mathrm{N}}{^{14}\mathrm{N}}{\mathrm{H}}^+
+ {^{15}\mathrm{N}}_2
 \mathrel{\mathop{\rightleftharpoons}^{k_5}_{k_{-5}}} 
{^{15}\mathrm{N}}_2{\mathrm{H}}^+
+ {^{15}\mathrm{N}}{^{14}\mathrm{N}},
\label{reaction_n2h_5}
\end{eqnarray}
\noindent
have common products. The two reaction pairs are related by
the $^{14}$N$\rightarrow {^{15}}$N substitution.
Using thermodynamic reasoning,
it is easy to verify %(with the help of the number densities)
that the following relationship
\begin{eqnarray}
  \frac{K_e^{(2)}}{K_e^{(3)}}
= \frac{K_e^{(5)}}{K_e^{(4)}}
= {K_e^{(6)}}
\label{keq_relations}
\end{eqnarray}
\noindent
is strictly fulfilled for $K_e^{(i)}=k_i/k_{-i}$,
where ${K_e^{(6)}}$ corresponds to reaction D6,
\begin{eqnarray}
{^{15}\mathrm{N}}{^{14}\mathrm{N}}{\mathrm{H}}^+
+ {^{15}\mathrm{N}}{^{14}\mathrm{N}}
 \mathrel{\mathop{\rightleftharpoons}^{k_6}_{k_{-6}}} 
{^{14}\mathrm{N}}{^{15}\mathrm{N}}{\mathrm{H}}^+
+ {^{14}\mathrm{N}}{^{15}\mathrm{N}}
.
\label{reaction_n2h_6} 
\end{eqnarray}
\noindent
Note that the factor $1/K_e^{(6)}$ also provides the thermal population
of $^{15}$N$^{14}$NH$^+$ relative to $^{14}$N$^{15}$NH$^+$.

%------------------------------------------------------------------------------------
\section{Discussion}
\label{sec:discussion}

The equilibrium constant $K_e$ for the proton transfer reaction 
\begin{eqnarray}
{\mathrm A} + {\mathrm{HB}}
\mathrel{\mathop{\rightleftharpoons}^{k_f}_{k_r}}
\mathrm{HA} +  {\mathrm B} 
\label{reaction_proton}
\end{eqnarray}
\noindent
under thermal equilibrium conditions is given by
\begin{eqnarray}
K_e = \frac{k_f}{k_r} = 
\frac{Q({\mathrm{HA}})}{Q({\mathrm{HB}})}
\frac{Q({\mathrm{B}})}{Q({\mathrm{A}})}
,
\label{equilibrium_constant_0}
\end{eqnarray}
\noindent
where $Q(\mathrm{X})$ is the full partition function for the species X.
Making the translation contribution explicit, 
we obtain  
\begin{eqnarray}
K_e = f_m^{3/2} \,
\frac{Q_{\mathrm{int}}({\mathrm{HA}})}{Q_{\mathrm{int}}({\mathrm{HB}})}
\frac{Q_{\mathrm{int}}({\mathrm{B}})}{Q_{\mathrm{int}}({\mathrm{A}})}
e^{\Delta E/k_{\mathrm{B}} T},
\label{equilibrium_constant}
\end{eqnarray}
\noindent
where the mass factor $f_m$ is given by
\begin{eqnarray}
f_m= \frac{m({\mathrm{HA}})\, m({\mathrm{B}})}{m({\mathrm{HB}})\, m({\mathrm{A}})}
\label{mass_factor}
\end{eqnarray}
\noindent
for $m$(X) denoting the mass of the species X, whereas $\Delta E$
stands for the zero-point energy difference between the reactants
and the products,
\begin{eqnarray}
\Delta E = E_0^{\mathrm{HA}} + E_0^{\mathrm{B}}
         - E_0^{\mathrm{HB}} - E_0^{\mathrm{A}} .
\label{delta_e}
\end{eqnarray}
\noindent
The zero-point energies $E_0$ are measured on an absolute energy scale.
For isotope fractionation reactions,
the internal partition function, $Q_{\mathrm{int}}$, includes only the rovibrational
degrees of freedom (no electronic contribution) and is given 
by the standard expression
\begin{eqnarray}
Q_{\mathrm{int}} = g \sum_J \sum_i \left(2J+1\right) e^{-\varepsilon_i^J/k_{\mathrm{B}} T},
\label{q_int} 
\end{eqnarray}
\noindent
where $\varepsilon_i^J = E_i^J-E_0^0$ for a total angular momentum $J$ 
is the rovibrational energy 
measured relative to the corresponding zero-point energy ($J$=0). 
The factor $(2J+1)$ accounts for the degeneracy relative to 
the space-fixed reference frame and
$g$ for the nuclear spin (hyperfine) degeneracy,
\begin{eqnarray}
g=\Pi_{\alpha} (2I_{N,\alpha}+1), 
\label{nuc_spin_g}
\end{eqnarray}
\noindent
in which $\alpha$ labels the constituent nuclei
having the nuclear spin $I_{N,\alpha}$.
For the nuclei considered in the present work, we have
$I_N($H$)=1/2$,
$I_N($D$)=1$,
$I_N(^{12}$C$)=0$,
$I_N(^{13}$C$)=1/2$,
$I_N(^{16}$O$)=0$,
$I_N(^{18}$O$)=0$,
$I_N(^{14}$N$)=1$, and
$I_N(^{15}$N$)=1/2$.
%%Note that the symmetry factor $\sigma$ sometimes used in the evaluation
%%of $K_e$ is of classical mechanical origin.

Introducing the ratio 
\begin{eqnarray}
R^{\mathrm{X}}_{\mathrm{Y}} & = & 
\frac{Q_{\mathrm{int}}({\mathrm{X}})}{Q_{\mathrm{int}}({\mathrm{Y}})},
\label{ratio_q}
\end{eqnarray}

\noindent
the equilibrium constant is compactly written as
\begin{eqnarray}
K_e  =  \frac{k_f}{k_r} 
= F_q \, e^{\Delta E/k_{\mathrm{B}} T},
\label{equilibrium_constant_compact}
\end{eqnarray}

\noindent
where the partition function factor $F_q$ is 
\begin{eqnarray}
F_q & = & f_m^{3/2} 
\, R^{\mathrm{HA}}_{\mathrm{HB}}
\, R^{\mathrm{B}}_{\mathrm{A}} .
\label{equilibrium_constant_factor}
\end{eqnarray}

\noindent
For reactions proceeding in the ground-rovibrational 
states of the reactants and the products,
the partition function ratios $R_{\mathrm{HB}}^{\mathrm{HA}}$ 
and $R_{\mathrm{A}}^{\mathrm{B}}$ are both equal to 1.
Even then the corresponding partition function factor $F_q$ of 
Eq. (\ref{equilibrium_constant_factor})
is, strictly speaking, different from 1 because of the mass term
$f_m$ defined by Eq. (\ref{mass_factor}).
For the reactions F1--F6 in Table \ref{table_kelvin}, for instance, 
the $f_m^{3/2}$ values are 0.998, 0.998, 0.997, 0.997, 0.995, 
and 1.002, respectively, which are different from 1
at most by 0.5\%. 
{\boldtext The $f_m^{3/2}$ values are computed from Eq. (\ref{mass_factor})
using the following atomic masses
$m(\mathrm{H})=1.007825035$,
$m(\mathrm{D})=2.014101779$,
$m(^{12}\mathrm{C})=12$,
$m(^{13}\mathrm{C})=13.003354826$,
$m(^{16}\mathrm{O})=15.99491463$,
and $m(^{18}\mathrm{O})=17.9991603$ u, as given by \cite{mills93}.
}

The terms of $Q_{\mathrm{int}}$ in Eq. (\ref{q_int})
decrease rapidly with energy and J.
In the low-temperature limit relevant for dark cloud conditions, 
the discrete rotational structure of the ground-vibrational state
provides the main contribution to $Q_{\mathrm{int}}$.
That said, the rotational energy cannot be treated 
as continuous and one must explicitly sum the terms to obtain  
$Q_{\mathrm{int}}$.  With increasing temperature, 
the rotational population in the ground-vibrational state increases
and other vibrational states may also become accessible, leading to    
partition function factors $F_q$, which may show (weak) 
temperature dependences.

{\boldtext
For a given potential energy surface, numerically exact full-dimenional strategies
insure the determination of accurate level energies and therefrom
accurate partition functions and equilibrium constants. 
To predict/estimate rate coefficients, we may use kinetic models, such as
e.g. the Langevin collision rate model for ion-molecule reactions.
Uncertainties in the rate coefficients are thus defined by uncertainties
in the model parameters.
In the case of the system CO+HCO$^+$, we employ the total rate coefficients from Table 3
of \cite{langer84} and the uncertainties of these quantities 
also provide the uncertainties of the rate coefficients derived in the present work.
}

%------------------------------------------------------------------------------------
\subsection{Reaction of CO with HCO$^+$}
\label{sec_discussion:hco+}

%####################
%\input{table_rate}
%--------------------------------------------------------------
\begin{table*}
\caption{\label{table_rate}
Equilibrium constants $K_e$,
partition function factors $F_q$, 
and rate coefficients $k_f, k_r$ (in 10$^{-10}$ cm$^3$s$^{-1}$) for 
the reactions of H/DCO$^+$ with CO.  }
\centering
\vspace{0.025cm}
\tiny 
\begin{tabular}{clrrrrrrrrr}
\hline\hline
 \noalign{\smallskip}
{Reaction} & {T (K)} 
& {5} & {10} & {20} & {40} & {60} & {80} & {100} & {200} & {300} \\
 \noalign{\smallskip}
\hline
 \noalign{\smallskip}
\small
%--------------------------------------------------------------
F1 %$^{13}$C$^{16}$O+H$^{12}$C$^{16}$O$^+$
            & $F_q$ &0.9854& 0.9832 & 0.9821 & 0.9816 & 0.9814 & 0.9813& 0.9813& 0.9815& 0.9824 \\ 
%$\rightleftharpoons$H$^{13}$C$^{16}$O$^+$+ $^{12}$C$^{16}$O
            & $K_e$   &34.64&  5.83&  2.39&  1.53&  1.32&  1.23&  1.17&  1.07&  1.04 \\  
%Label~F1  % reaction (\ref{reaction2})
            & $K_e$(see $a$)   &     & 6.69&     &     &     &     &  1.19 &  1.08 &  1.05 \\
            & $K_e$(see $b$)   & 6.0 & 2.5 & 1.6 & 1.3 & 1.2 & 1.1 &   1.1 &   1.0 &   1.0 \\
         & $(k_f,k_r)$ &(9.33,0.27)&(7.85,1.35)&(6.35,2.65)&(5.20,3.40)&(4.67,3.53)&(4.30,3.50)&(4.05,3.45)
                       &(3.21,2.99)&(2.65,2.55) \\  
       & $(k_f,k_r)^b$ &(8.2,1.4)&(6.5,2.7)&(5.5,3.5)&(4.8,3.8)&(4.4,3.8)&(4.1,3.7)&(3.9,3.6)
                       &(3.2,3.0)&(2.6,2.6) \\  
         &$(k_f,k_r)^c$&  &     &     &     &     &(4.2,3.6)& &(3.2,3.0)&(2.6,2.5) \\
 \noalign{\smallskip}
F2 %$^{13}$C$^{18}$O+H$^{12}$C$^{18}$O$^+$
            & $F_q$ &0.9853& 0.9831 & 0.9821 & 0.9816 & 0.9814 & 0.9813& 0.9813& 0.9814& 0.9822 \\ 
%$\rightleftharpoons$H$^{13}$C$^{18}$O$^+$+ $^{12}$C$^{18}$O
            & $K_e$   &34.93&  5.85&  2.40&  1.53&  1.32&  1.23&  1.17&  1.07&  1.04 \\  
%Label~F2 
            & $K_e$(see $a$)   &     & 6.65&     &     &     &     &  1.19 &  1.08 &  1.05 \\
            & $K_e$(see $b$)   & 6.0 & 2.5 & 1.6 & 1.3 & 1.2 & 1.1 &   1.1 &   1.0 &   1.0 \\
         & $(k_f,k_r)$ &(9.3,0.27)&(7.86,1.34)&(6.35,2.65)&(5.21,3.39)&(4.67,3.53)&(4.30,3.50)&(4.05,3.45)
                       &(3.21,2.99)&(2.65,2.55) \\  
       & $(k_f,k_r)^b$ &(8.2,1.4)&(6.5,2.7)&(5.5,3.5)&(4.8,3.8)&(4.4,3.8)&(4.1,3.7)&(3.9,3.6)
                       &(3.2,3.0)&(2.6,2.6) \\  
 \noalign{\smallskip}
F3 %$^{12}$C$^{18}$O+H$^{12}$C$^{16}$O$^+$
            & $F_q$ & 0.9964& 0.9951 & 0.9945& 0.9942& 0.9941& 0.9941& 0.9941& 0.9941 & 0.9946\\
%$\rightleftharpoons$H$^{12}$C$^{18}$O$^+$+ $^{12}$C$^{16}$O
            & $K_e$    &  3.59&  1.89&  1.37&  1.17&  1.11&  1.08&  1.06&  1.03&  1.02 \\  
%Label~F3 % reaction (\ref{reaction3})
            &$K_e$(see $b$) & 16.4 & 4.1 & 2.0 & 1.4 & 1.3 & 1.2 & 1.2 & 1.1 &  1.0 \\
         & $(k_f,k_r)$ &(7.51,2.09)&(6.01,3.19)&(5.20,3.80)&(4.63,3.97)&(4.31,3.89)&(4.04,3.76)&(3.86,3.64)
                       &(3.14,3.06)&(2.62,2.58) \\  
       & $(k_f,k_r)^b$ &(9.0,0.6)&(7.4,1.8)&(6.0,3.0)&(5.0,3.6)&(4.6,3.6)&(4.2,3.6)&(4.0,3.5)
                       &(3.2,3.0)&(2.7,2.5) \\  
         &$(k_f,k_r)^c$&  &     &     &     &     &(4.2,3.5)& &(3.2,2.9)&(2.7,2.6) \\
 \noalign{\smallskip}
F4 %$^{13}$C$^{18}$O+H$^{13}$C$^{16}$O$^+$
            & $F_q$ & 0.9963& 0.9951 & 0.9945& 0.9942& 0.9941& 0.9941& 0.9940& 0.9941 & 0.9943\\
%$\rightleftharpoons$H$^{13}$C$^{18}$O$^+$+ $^{13}$C$^{16}$O
            & $K_e$    &  3.61&  1.90&  1.37&  1.17&  1.11&  1.08&  1.06&  1.03&  1.02 \\  
%Label~F4 % reaction (\ref{reaction3})
            &$K_e$(see $b$) & 16.4 & 4.1 & 2.0 & 1.4 & 1.3 & 1.2 & 1.2 & 1.1 &  1.0 \\
         & $(k_f,k_r)$ &(7.52,2.08)&(6.02,3.18)&(5.21,3.79)&(4.63,3.97)&(4.31,3.89)&(4.05,3.75)&(3.86,3.64)
                       &(3.14,3.06)&(2.62,2.58) \\  
       & $(k_f,k_r)^b$ &(9.0,0.6)&(7.4,1.8)&(6.0,3.0)&(5.0,3.6)&(4.6,3.6)&(4.2,3.6)&(4.0,3.5)
                       &(3.2,3.0)&(2.7,2.5) \\  
 \noalign{\smallskip}
F5 %$^{13}$C$^{18}$O+H$^{12}$C$^{16}$O$^+$
            & $F_q$ & 0.9818& 0.9783& 0.9767& 0.9759& 0.9756& 0.9755& 0.9754& 0.9756& 0.9769\\
%$\rightleftharpoons$H$^{13}$C$^{18}$O$^+$+ $^{12}$C$^{16}$O
            & $K_e$   &125.26& 11.05&  3.28&  1.79&  1.46&  1.32&  1.24&  1.10&  1.06 \\  
%Label~F5 %reaction (\ref{reaction_isto})
            &$K_e$(see $a$) &      &13.37&     &     &     &     & 1.27& 1.11&  1.06\\
            & $K_e$(see $b$) & 81.5 & 9.0 & 3.0 & 1.7 & 1.4 & 1.3 & 1.2 & 1.1 & 1.1 \\
         & $(k_f,k_r)$ &(9.52,0.08)&(8.44,0.76)&(6.90,2.10)&(5.52,3.08)&(4.87,3.33)&(4.44,3.36)&(4.16,3.34)
                       &(3.25,2.95)&(2.67,2.53) \\  
       & $(k_f,k_r)^b$ &(9.5,0.1)&(8.3,0.9)&(6.8,2.2)&(5.5,3.1)&(4.8,3.4)&(4.4,3.4)&(4.2,3.3)
                       &(3.3,2.9)&(2.7,2.5) \\  
 \noalign{\smallskip}
%%reaction (\ref{reaction_diff})
F6 % $^{13}$C$^{16}$O+H$^{12}$C$^{18}$O$^+$
            & $F_q$ & 0.9890& 0.9880 & 0.9875& 0.9873& 0.9872& 0.9872& 0.9872 & 0.9873& 0.9878\\
%$\rightleftharpoons$H$^{13}$C$^{16}$O$^+$+ $^{12}$C$^{18}$O
            & $K_e$   &  9.66&  3.09&  1.75&  1.31&  1.19&  1.14&  1.11&  1.04&  1.03 \\  
%Label~F6  % reaction (\ref{reaction_diff})
            &$K_e$(see $a$)  &      & 3.33 &      &      &      &      & 1.11& 1.05&  1.03\\
            & $K_e$(see $b$) &  0.37 & 0.61 & 0.78 & 0.88 & 0.92 & 0.94 & 0.95 & 0.98 & 0.98 \\
         & $(k_f,k_r)$ &(8.70,0.90)&(6.95,2.25)&(5.72,3.28)&(4.88,3.72)&(4.46,3.74)&(4.15,3.65)&(3.94,3.56)
                       &(3.17,3.03)&(2.63,2.57) \\  
       & $(k_r,k_f)^b$ &(3.0,6.6)&(3.7,5.5)&(4.1,4.9)&(4.1,4.5)&(4.0,4.2)&(3.8,4.0)&(3.7,3.8)
                       &(3.1,3.1)&(2.6,2.6) \\  
          &$(k_f,k_r)^c$&  &     &     &     &     &(4.3,4.1)& &(3.1,3.2)&(2.6,2.5) \\
%--------------------------------------------------------------
 \noalign{\smallskip}
%%D variants & \\
\cline{2-11}
 \noalign{\smallskip}
%reaction (\ref{reaction2})
F1(D) %$_{\mathrm{d}}$ %$^{13}$C$^{16}$O+D$^{12}$C$^{16}$O$^+$
            & $F_q$ &0.9764& 0.9733 & 0.9718 & 0.9710 & 0.9708 & 0.9707& 0.9706& 0.9716& 0.9738 \\ 
%$\rightleftharpoons$D$^{13}$C$^{16}$O$^+$+ $^{12}$C$^{16}$O
            & $K_e$   &74.64&  8.51&  2.87&  1.67&  1.39&  1.27&  1.21&  1.08&  1.05 \\  
%Label~F1$_{\mathrm{d}}$
   & $K_e$(see $a$)   &     &  9.78&      &      &      &      &  1.22&  1.09&  1.05 \\  
         & $(k_f,k_r)$ &(9.47,0.13)&(8.23,0.97)&(6.68,2.32)&(5.38,3.22)&(4.77,3.43)&(4.37,3.43)&(4.10,3.40)
                       &(3.22,2.98)&(2.66,2.54) \\  
 \noalign{\smallskip}
%---------------
F2(D) %$_{\mathrm{d}}$ %$^{13}$C$^{18}$O+D$^{12}$C$^{18}$O$^+$
            & $F_q$ &0.9762& 0.9731 & 0.9716 & 0.9709 & 0.9707 & 0.9706& 0.9705& 0.9715& 0.9738 \\ 
%$\rightleftharpoons$D$^{13}$C$^{18}$O$^+$+ $^{12}$C$^{18}$O
            & $K_e$   &75.06&  8.53&  2.88&  1.67&  1.39&  1.27&  1.21&  1.08&  1.05 \\  
%Label~F2$_{\mathrm{d}}$
         & $(k_f,k_r)$ &(9.47,0.13)&(8.23,0.97)&(6.68,2.32)&(5.38,3.22)&(4.77,3.43)&(4.37,3.43)&(4.10,3.40)
                       &(3.22,2.98)&(2.66,2.54) \\  
 \noalign{\smallskip}
%-------------
%reaction (\ref{reaction3})
F3(D) %$_{\mathrm{d}}$ %$^{12}$C$^{18}$O+D$^{12}$C$^{16}$O$^+$
            & $F_q$ & 0.9940& 0.9920 & 0.9911& 0.9906& 0.9904& 0.9904& 0.9903& 0.9905 & 0.9911\\
% $\rightleftharpoons$D$^{12}$C$^{18}$O$^+$+ $^{12}$C$^{16}$O
            & $K_e$    &  4.65&  2.14&  1.46&  1.20&  1.13&  1.09&  1.07&  1.03&  1.02 \\  
%Label~F3$_{\mathrm{d}}$
         & $(k_f,k_r)$ &(7.90,1.70)&(6.27,2.93)&(5.34,3.66)&(4.69,3.91)&(4.34,3.86)&(4.07,3.73)&(3.88,3.62)
                       &(3.14,3.06)&(2.62,2.58) \\  
 \noalign{\smallskip}
%-------------
F4(D) %$_{\mathrm{d}}$ % $^{13}$C$^{18}$O+D$^{13}$C$^{16}$O$^+$
            & $F_q$ & 0.9938& 0.9918 & 0.9909& 0.9905& 0.9904& 0.9903& 0.9902& 0.9905 & 0.9911\\
%$\rightleftharpoons$D$^{13}$C$^{18}$O$^+$+ $^{13}$C$^{16}$O
            & $K_e$    &  4.67&  2.15&  1.46&  1.20&  1.13&  1.09&  1.07&  1.03&  1.02 \\  
%Label~F4$_{\mathrm{d}}$
         & $(k_f,k_r)$ &(7.91,1.69)&(6.28,2.92)&(5.34,3.66)&(4.69,3.91)&(4.34,3.86)&(4.07,3.73)&(3.88,3.62)
                       &(3.15,3.05)&(2.62,2.58) \\  
 \noalign{\smallskip}
%reaction (\ref{reaction_isto})
F5(D) %$_{\mathrm{d}}$ %$^{13}$C$^{18}$O+D$^{12}$C$^{16}$O$^+$
            & $F_q$ & 0.9704& 0.9653& 0.9630& 0.9618& 0.9614& 0.9612& 0.9611& 0.9623& 0.9651\\
%$\rightleftharpoons$D$^{13}$C$^{18}$O$^+$+ $^{12}$C$^{16}$O
            & $K_e$   &348.81& 18.30&  4.19&  2.00&  1.57&  1.39&  1.29&  1.11&  1.06 \\  
%Label~F5$_{\mathrm{d}}$
         & $(k_f,k_r)$ &(9.57,0.03)&(8.72,0.48)&(7.27,1.73)&(5.74,2.86)&(5.01,3.19)&(4.53,3.27)&(4.22,3.28)
                       &(3.27,2.93)&(2.68,2.52) \\  
 \noalign{\smallskip}
%reaction (\ref{reaction_diff})
F6(D) %$_{\mathrm{d}}$ % $^{13}$C$^{16}$O+D$^{12}$C$^{18}$O$^+$
            & $F_q$ & 0.9823& 0.9811 & 0.9805& 0.9803& 0.9802& 0.9801& 0.9801 & 0.9808& 0.9826\\
%$\rightleftharpoons$D$^{13}$C$^{16}$O$^+$+ $^{12}$C$^{18}$O
            & $K_e$   & 16.06&  3.97&  1.97&  1.39&  1.24&  1.17&  1.13&  1.05&  1.03 \\  
%Label~F6$_{\mathrm{d}}$
%           & $k_f$ & 9.0   & 7.3 & 6.0 & 5.0 & 4.5& 4.2 & 4.0& 3.2& 2.6 \\
%           & $k_r$ & 0.6   & 1.8 & 3.0 & 3.6 & 3.7& 3.6 & 3.5& 3.0& 2.6 \\
         & $(k_f,k_r)$ &(9.04,0.56)&(7.35,1.85)&(5.97,3.03)&(5.00,3.60)&(4.53,3.67)&(4.20,3.60)&(3.97,3.53)
                       &(3.18,3.02)&(2.64,2.56) \\  
 \noalign{\smallskip}
\hline
%--------------------------------------------------------------
\end{tabular}
\tablefoot{
\tablefoottext{a}{\cite{lohr98}}
\tablefoottext{b}{\cite{langer84}}
\tablefoottext{c}{\cite{smith80a}}
}
\end{table*}
%\end{sidewaystable}
%}
%-----------------------------------------------------------------------

%####################

The equilibrium constants for HCO$^+$ reacting with CO are given 
in Table \ref{table_rate}.  Our $K_e$ values are obtained 
in accordance with Eq. (\ref{equilibrium_constant_compact}) 
by direct evaluation of the internal partition functions $Q_{\mathrm{int}}$
from the computed rovibrational energies. 
The forward reaction $k_f$ 
and backward reaction $k_r$ rate coefficients 
are calculated using our $\Delta E$ values
and the total temperature-dependent 
rate coefficients $k_{\mathrm{T}}$ given by \cite{langer84},
where 
\begin{eqnarray}
k_{\mathrm{T}} = k_f+k_r ,
\label{rate_langer}
\end{eqnarray}

\noindent
such that
\begin{eqnarray}
k_f & = & k_{\mathrm{T}} K_e/(K_e+1), %%\frac{K}{K+1}
\label{kf_langer}\\
k_r & = & k_{\mathrm{T}} /(K_e+1) . %%\frac{1}{K+1}
\label{kr_langer}
\end{eqnarray}
\noindent 
The results for the deuterium variants 
%of reactions (\ref{reaction_diff})-(\ref{reaction_isto})
are also listed in Table \ref{table_rate}.
Their rate coefficients $k_f$ and $k_r$ are calculated
assuming the same total rate coefficients $k_{\mathrm{T}}$
as for the H-containing forms (due to nearly equal
reduced masses).
For the purpose of comparison, note that the Langevin
rate for CO+HCO$^+$ is
$k_{\mathrm{L}} = 8.67 \times 10^{-10}$ cm$^3$s$^{-1}$.

The partition function factors $F_q$ 
deviate from 1 by approximately 2\% in Table \ref{table_rate}. 
They also exhibit mar\-gi\-nal temperature dependences. 
This reflects the influence of rotational and vibrational
excitations in the reactants and the products.
Only the rotationally excited ground-vibrational states 
contribute to $Q_{\mathrm{int}}$
at temperatures $T < 200$ K.
%For the HCO$^+$ isotopologues.
The contribution of the bending $\nu_2$ level is 0.5\% at 200 K
and 3.6--3.8\% at 300 K, whereas the contributions from    
2$\nu_2$ are 0.1\% at 300 K.
To appreciate the effect of $F_q$, 
we employed the rate coefficients measured at 80 K by \cite{smith80a} 
to determine the $\Delta E$ value 
for reactions F1, F3, and F6
by means of Eq. (\ref{equilibrium_constant_compact}).
Using the $F_q$ values from Table \ref{table_rate},
we obtain $\Delta E/k_{\mathrm{B}}$ of 13.8, 15.1, and 4.8 K, respectively.
For $F_q=1$,
we find 12.3, 14.6, and 3.8 K, which are lower by
1.5 K (12\%), 0.5 K (3\%), and 1 K (26\%)
than the former $F_q \ne 1$ results.

The equilibrium constants $K_e$ reported by \cite{langer84} deviate from the present
results and those of \cite{lohr98} very prominently at low temperatures
in Table \ref{table_rate}.
At 10 K, we see deviations of 
43\% and 217\% with respect to our values for 
reactions F1 and F3, respectively, and
the related rate coefficients $k_f, k_r$ are accordingly different.
An even larger discrepancy is seen for 
reaction F6 of Eq. (\ref{reaction_diff}),
which was previously predicted to be endothermic.
In accordance with this, the values of $k_f$ and $k_r$ 
derived by \cite{langer84} are given in the reverse positions
as $(k_r,k_f)$ for reaction F6 in Table \ref{table_rate}. 

The deuterium variants in Table \ref{table_rate} are associated with slightly lower
$F_q$ values and somewhat higher low-temperature $K_e$, resulting in
somewhat faster foward reactions and slower backward reactions.
 
%-------------------------------------------------------
\subsection{Astrochemical implications}
\label{discussion:astro}

%#############################
%\input{table_evelyne_final}
%---------------------- table Dis ----------------------------------
\begin{table*}
\caption{\label{tab:result}
Isotopic fractionation ratios at 10 K
for three H$_2$ densities, $n$(H$_2$).  }
\centering
\vspace{0.025cm}
%\tiny 
\begin{tabular}{crrrcrrr}
\hline\hline
 \noalign{\smallskip}
 & & \multicolumn{2}{c}{Model A: LGFA84}
 & & \multicolumn{2}{c}{Model B: Present results}
\\
 \noalign{\smallskip}
\cline{3-4} \cline{6-7}
 \noalign{\smallskip}
{$n$(H$_2$) (cm$^{-3}$)} & {Species} & {$x^{a,b}$} & {$R_{\mathrm{A}}$}
&& {$x^{a,b}$} & {$R_{\mathrm{B}}$} & $\delta$ (\%)
\\
 \noalign{\smallskip}
\hline
 \noalign{\smallskip}
%\small
 \noalign{\smallskip}
$10^4$ & $^{12}$C$^{16}$O & 6.9290(-5) &       &~~&  6.9283(-5) &         \\
       & $^{13}$C$^{16}$O & 1.1534(-6) &  60.1  & &  1.1518(-6) &  60.1  \\
       & $^{12}$C$^{18}$O & 1.2232(-7) &  566   & &  1.3095(-7) &  529   & 6.5 \\
       & $^{13}$C$^{18}$O & 2.1278(-9) & 32\,564  & &   2.2849(-9)& 30\,322  & 6.9 \\
 \noalign{\smallskip}
     & H$^{12}$C$^{16}$O$^+$ & 7.7399(-9) &        & &   7.7204(-9)  & \\
     & H$^{13}$C$^{16}$O$^+$ & 1.5393(-10)&  50.3  & &   1.7448(-10) &  44.2  & 12.0\\
     & H$^{12}$C$^{18}$O$^+$ & 1.8001(-11)&  430   & &   1.6805(-11) &  459   & -6.7\\
     & H$^{13}$C$^{18}$O$^+$ & 3.3928(-13)&  22\,812 & &   3.6906(-13) & 20\,919  & 8.3 \\
     & D$^{12}$C$^{16}$O$^+$ & 2.0767(-10)&        & &   2.0613(-10) & \\
     & D$^{13}$C$^{16}$O$^+$ & 3.5964(-12)& 57.7   & &   5.1283(-12) & 40.2   & 30.4 \\
     & $e^-$                 & 3.8096(-8) &        & &   3.8096(-8)  & \\
 \noalign{\smallskip}
$10^5$ & $^{12}$C$^{16}$O & 6.9704(-5) &        & &  6.9686(-5) &         \\
       & $^{13}$C$^{16}$O & 1.1604(-6) &  60.1  & &  1.1585(-6) &  60.2  \\
       & $^{12}$C$^{18}$O & 1.0029(-7) &  695   & &  1.2071(-7) &  577   & 17.0\\
       & $^{13}$C$^{18}$O & 1.8244(-9) & 38\,207  & &  2.2083(-9) & 31\,556  & 17.4\\
 \noalign{\smallskip}
     & H$^{12}$C$^{16}$O$^+$ & 3.0114(-9) &        & &   2.9847(-9)  & \\
     & H$^{13}$C$^{16}$O$^+$ & 7.6967(-11)&  39.1  & &   1.0465(-10) &  28.5  & 27.1\\
     & H$^{12}$C$^{18}$O$^+$ & 8.1969(-12)&  367   & &   7.0938(-12) &  421   &-14.7\\
     & H$^{13}$C$^{18}$O$^+$ & 1.8653(-13)&  16\,144 & &   2.3351(-13) & 12\,782  & 20.8\\
     & D$^{12}$C$^{16}$O$^+$ & 7.4521(-11)&        & &   7.2876(-11) & \\
     & D$^{13}$C$^{16}$O$^+$ & 1.3020(-12)&  57.2  & &   2.9330(-12) & 24.8   & 56.6 \\
     & $e^-$                 & 1.0383(-8) &        & &   1.0383(-8)  & \\
 \noalign{\smallskip}
$10^6$ & $^{12}$C$^{16}$O & 6.9836(-5) &        & &  6.9836(-5) &         \\
       & $^{13}$C$^{16}$O & 1.1632(-6) &  60.0  & &  1.1614(-6) &  60.1  \\
       & $^{12}$C$^{18}$O & 7.8707(-8) &  887   & &  1.1045(-7) &  632   & 28.7\\
       & $^{13}$C$^{18}$O & 1.4089(-9) & 49\,568  & &  1.9983(-9) & 34\,948  & 29.5\\
 \noalign{\smallskip}
     & H$^{12}$C$^{16}$O$^+$ & 1.0193(-9) &        & &   9.9657(-10) & \\
     & H$^{13}$C$^{16}$O$^+$ & 3.3343(-11)&  30.6  & &   5.6676(-11) &  17.6  & 42.5\\
     & H$^{12}$C$^{18}$O$^+$ & 3.1755(-12)&  321   & &   2.5647(-12) &  389   &-21.1\\
     & H$^{13}$C$^{18}$O$^+$ & 8.7690(-14)&  11\,624 & &   1.3324(-13) &  7\,480  & 35.7\\
     & D$^{12}$C$^{16}$O$^+$ & 2.4521(-11)&        & &   2.3308(-11) & \\
     & D$^{13}$C$^{16}$O$^+$ & 4.2140(-13)&  58.2  & &   1.6270(-12) & 14.3   & 75.4 \\
     & $e^-$                 & 3.0251(-9) &        & &   3.0251(-9)  & \\
 \noalign{\smallskip}
\hline
%-------------------------------------------------------
\end{tabular}
\tablefoot{
\tablefoottext{a}{Fractional abundances $x$ are given relative to H$_2$.}
\tablefoottext{b}{Numbers in parentheses denote powers of 10.}
}
\end{table*}
%-----------------------------------------------------------------------

%#############################

We investigated the role of these new derived exothermicities 
under different density conditions relevant 
to cold dark interstellar clouds.
We display in Table \ref{tab:result} steady-state results for
isotopic ratios of CO, HCO$^+$ and DCO$^+$ 
for two chemical models performed at a temperature of 10 K with a cosmic ionization 
rate $\zeta$ of 1.3 $\times$ 10$^{-17}$ s$^{-1}$ per H$_2$ molecule 
with the old $\Delta E$ values by Langer et al. (Model A: LGFA84) 
and the present $\Delta E$ values 
listed in Table \ref{table_kelvin} (Model B).
The ratios of the principal isotope to the minor isotope
obtained for Model A, $R_{\mathrm{A}}$, and for Model B, $R_{\mathrm{B}}$,
are compared using the relative difference $\delta$,
\begin{eqnarray}
\delta = 1-{R_{\mathrm{B}}}/{R_{\mathrm{A}}}.
\end{eqnarray} 

The chemical network contains 288 chemical species including $^{13}$C and $^{18}$O 
containing molecules 
as well as deuterated species and more than 5\,000 reactions. 
We assumed that the elemental $^{12}$C/$^{13}$C and $^{16}$O/$^{18}$O isotopic ratios 
are 60 and 500, 
so that any deviation relative to these values measures 
the amount of enrichment/depletion with respect to the elemental ratios.  
For the $^{13}$C$^{18}$O-containing molecules the value of
30\,000 is the reference.
{\boldtext
%%The zero-point energies of other isotopically substitutes do not pose any problem 
The zero-point energies of other isotopic substitutes do not pose any problem 
because the reactions involved in the interstellar chemical networks are significantly 
exothermic and the solutions of the chemical equations 
are independent of these quantities.
}

The isotopic fractionation reactions are introduced explicitly in the chemical network, 
whereas the other reactions 
involving isotopologues are built automatically from the reactions involving 
the main isotope in the chemical code. 
The adopted   method has first been presented in \cite{lebourlot93}, 
where statistical arguments were used to derive the various branching ratios in the chemical reactions. 
The procedure is limited to three carbon-containing molecules 
(oxygen-containing molecules have a maximum of two oxygen atoms in our chemical network) 
and does not disitinguish between C$^{13}$CC- or $^{13}$CCC-containing species. 
A similar approach has recently been applied by \cite{roellig13} 
for photon-dominated region models. However, \cite{roellig13} used 
the old (LGFA84) exothermicity values.
We also explicitly introduce the relation given by \cite{langer84} that 
$k_f + k_r = k_{\mathrm{T}}$. 
The forward reaction $k_f$ and reverse reaction $k_r$ rate coefficients 
involved in the isotopic exchange reaction are then evaluated 
from the total rate coefficient $k_{\mathrm{T}}$ as follows
\begin{eqnarray}
k_f = k_{\mathrm{T}} \frac{1}{1+\exp(-\Delta E/k_{\mathrm{B}} T)}
\label{evelyne1}
\end{eqnarray}

\noindent
and
\begin{eqnarray} 
k_r = k_{\mathrm{T}} \frac{\exp(-\Delta E/k_{\mathrm{B}} T)}{1+\exp(-\Delta E/k_{\mathrm{B}} T)}
.
\label{evelyne2}
\end{eqnarray}

\noindent 
These expressions have also been included in the study of fractionation in diffuse 
clouds presented by \cite{liszt07}.
 
The results summarized in Table \ref{tab:result} show that CO/$^{13}$CO has 
the elemental value, 
whereas rarer isotopologues are very slightly depleted. 
The results for Models A and B are also very similar because no differences 
were used 
for the reaction rate coefficients between $^{13}$C$^+$ and CO. 
However, more significant are the differences for the results for 
the isotopic ratio of HCO$^+$, 
which directly arise from the variations of the exothermicities found in the present work. 
We also introduced a fractionation reaction for the deuterated isotope, 
whose rotational frequencies have been measured in the laboratory \citep{caselli05} 
and are detected in the interstellar medium \citep{guelin82,caselli02}.
%and awaits still an astrophysical detection. 
As the exothermicity of the deuterated isotopologues is somewhat higher, 
the isotopic $^{13}$C ratio is somewhat lower than in the hydrogenic counterpart. 
%% which helps the detection of the D$^{13}$CO$^+$ molecular ion. 

%The key message of Table \ref{tab:result} is that the new exothermicities
The general trend seen in Table \ref{tab:result} 
is that the new Model B 
predicts lower fractional abundances $x$ 
for H$^{12}$C$^{16}$O$^+$ (up to 2\%)
and D$^{12}$C$^{16}$O$^+$ (up to 5\%), 
lower relative abundances $R_{\mathrm{B}}$ of H$^{12}$C$^{18}$O$^+$ (7--21\%),  
and higher relative abundances of the $^{13}$C-containing
isotopologues (up to 40\% for the hydrogenic forms and up to 75\% 
for the deuterated forms)
than Model A.

%---------------------------------------------------------
\subsection{Reaction of N$_2$ with N$_2$H$^+$}
\label{sec_discussion:nnh+}

Molecular nitrogen is a homonuclear diatomic molecule 
with a $X {^1}\Sigma_g^+$ ground-electronic state 
with the three naturally occurring isotopologues: 
$^{14}$N$_2$, $^{14}$N$^{15}$N, and $^{15}$N$_2$.
Whereas $^{14}$N is a spin-1 boson, 
$^{15}$N is a spin-1/2 fermion, such that 
the two symmetric forms $^{14}$N$_2$ and $^{15}$N$_2$ follow different
nuclear spin statistics. 
In the states with a higher nuclear spin degeneracy (ortho states), 
we have $g=(I_N+1)(2 I_N + 1)$,
whereas $g=I_N(2 I_N + 1)$ holds 
for the states with lower nuclear spin degeneracy (para states).
To properly account for this effect,
we evaluated the internal partition functions separately for
even and odd $J$ values,
\begin{eqnarray}
Q_{\mathrm{\mathrm{evenJ}}} & = & {\sum_{J=0}}' \sum_i (2J+1) \, e^{-\varepsilon_i^J/k_{\mathrm{B}} T}
,
\\
Q_{\mathrm{\mathrm{oddJ}}} & = & {\sum_{J=1}}' \sum_i (2J+1) \, e^{-\varepsilon_i^J/k_{\mathrm{B}} T}
,
~~~
\end{eqnarray}

\noindent
where $\Sigma'$ denotes summation in steps of 2.
Multiplying each term by the appropriate nuclear spin (hyperfine) 
degeneracy factor,
we obtain the partition function for N$_2$ as
% the two contributions can be combined as
\begin{eqnarray}
Q_{\mathrm{int}}(^{14}\mathrm{N}_2) & = & 6 Q_{\mathrm{evenJ}}
                               + 3 Q_{\mathrm{oddJ}}
,
\label{n14} \\
Q_{\mathrm{int}}(^{15}\mathrm{N}_2) & = & 3 Q_{\mathrm{oddJ}}
                               +  Q_{\mathrm{evenJ}}
.
\label{n15}
\end{eqnarray}
\noindent
%In other words, 
%the degeneracy of the even rotational levels of $^{14}$N$_2$
%is effectively twice that of the odd levels, whereas 
%the degeneracy of the odd rotational levels of $^{15}$N$_2$
%is three times that of the even levels. 
%
$^{14}$N$^{15}$N is not a homonuclear diatomic molecule, such that
\begin{eqnarray}
Q_{\mathrm{int}}(^{14}\mathrm{N}^{15}\mathrm{N}) 
& = & 6 \left( Q_{\mathrm{evenJ}} + Q_{\mathrm{oddJ}} \right)
.
\end{eqnarray}

%#######################
%\input{table_rate_n2h}
%--------------------------------------------------------------
\begin{table*}
\caption{\label{table_rate_n2h}
Equilibrium constants $K_e$,
partition function factors $F_q$, 
and rate coefficients $k_f, k_r$ 
(in 10$^{-10}$ cm$^3$s$^{-1}$) for 
the reactions of N$_2$H$^+$ with N$_2$.  }
\centering
\vspace{0.025cm}
\tiny 
\begin{tabular}{clrrrrrrrrr}
\hline\hline
 \noalign{\smallskip}
{Reaction} & {T (K)} 
& {5} & {10} & {20} & {40} & {60} & {80} & {100} & {200} & {292}
\\
 \noalign{\smallskip}
\hline
 \noalign{\smallskip}
\small
%--------------------------------------------------------------
D1 %$^{14}$N$_2$H$^+$+ $^{15}$N$_2$
            & $F_q$ &1.0561& 0.9858 & 0.9834 & 0.9828 & 0.9826 & 0.9825& 0.9825& 0.9829& 0.9840 \\ 
%$\rightleftharpoons ^{15}$N$_2$H$^+$+ $^{14}$N$_2$
            & $K_e^{(1)}$   &12.74&  3.42&  1.83&  1.34&  1.21&  1.15&  1.11&  1.05&  1.03 \\  
   & $(k_1,k_{-1})$ &(7.52,0.59)&(6.28,1.83)&(5.25,2.86)&(4.65,3.46)&(4.44,3.67)&  (4.33,3.78)&(4.27,3.84)&(4.15,3.96)&  (4.11,4.00) \\
        & $(k_f,k_r)^a$ &     &     &     &     &     &  (4.8,4.1)&     &     &  (4.1,4.1) \\
 \noalign{\smallskip}
\cline{2-11}
 \noalign{\smallskip}
D2 % $^{14}$N$_2$H$^+$+ $^{15}$N$^{14}$N
            & $F_q$ & 0.5101& 0.4941 & 0.4934& 0.4932& 0.4931& 0.4931& 0.4931& 0.4932 & 0.4937\\
% $\rightleftharpoons ^{14}$N$^{15}$NH$^+$+ $^{14}$N$_2$
            & $K_e^{(2)}$    &  4.05&  1.39&  0.83&  0.64&  0.59&  0.56&  0.55&  0.52&  0.51 \\  
   & $(k_2,k_{-2})$ & (6.50,1.61)&(4.72,3.39)&(3.67,4.44) &(3.16,4.95) &(3.00,5.11)&(2.92,5.19)&(2.87,5.24)&(2.77,5.34)&(2.74,5.36) \\
 \noalign{\smallskip}
D3 %$^{14}$N$_2$H$^+$+ $^{14}$N$^{15}$N
            & $F_q$ & 0.5147& 0.4989& 0.4983& 0.4982& 0.4982& 0.4982& 0.4982& 0.4982& 0.4983\\
%$\rightleftharpoons ^{15}$N$^{14}$NH$^+$+ $^{14}$N$_2$
            & $K_e^{(3)}$   &0.78  & 0.55 &  0.52&  0.52&  0.51&  0.51&  0.51&  0.50&  0.50 \\  
& $(k_3,k_{-3})$ &(3.54,4.57)& (3.08,5.03)& (2.89,5.22)&(2.79,5.32)&(2.76,5.35)&  (2.74,5.37)& (2.73,5.38)& (2.72,5.39)&  (2.71,5.40) \\
 \noalign{\smallskip}
\cline{2-2}
 \noalign{\smallskip}
& $(k_{23},k_{-23})$ &(10.05,2.08)& (7.80,3.89)& (6.56,4.75)&(5.95,5.12)&(5.76,5.22)&  (5.66,5.28)& (5.60,5.31)& (5.49,5.37)&  (5.45,5.38) \\
& $(k_f,k_r)^a$ &     &     &     &     &     &  (4.6,4.1)&     &     &  (4.1,4.1) \\
 \noalign{\smallskip}
\cline{2-11}
 \noalign{\smallskip}
D4 %$^{14}$N$^{15}$NH$^+$+ $^{15}$N$_2$
            & $F_q$ & 2.0703& 1.9952 & 1.9933& 1.9929& 1.9927& 1.9927& 1.9926 & 1.9928& 1.9932\\
%$\rightleftharpoons ^{15}$N$_2$H$^+$+ $^{14}$N$^{15}$N
            & $K_e^{(4)}$   &  3.15&  2.46&  2.21&  2.10&  2.06&  2.05&  2.03&  2.01&  2.01 \\  
& $(k_{4},k_{-4})$ &(6.16,1.95)& (5.77,2.34)& (5.59,2.52)&(5.49,2.62)&(5.46,2.65)&  (5.45,2.66)& (5.44,2.67)& (5.42,2.69)&  (5.41,2.70) \\
 \noalign{\smallskip}
D5 % $^{15}$N$^{14}$NH$^+$+ $^{15}$N$_2$
            & $F_q$ & 2.0521& 1.9761 & 1.9734& 1.9727& 1.9724& 1.9723& 1.9722 & 1.9730& 1.9748\\
% $\rightleftharpoons ^{15}$N$_2$H$^+$+ $^{14}$N$^{15}$N
            & $K_e^{(5)}$   & 16.42&  5.59&  3.32&  2.56&  2.35&  2.25&  2.19&  2.08&  2.05 \\  
& $(k_{5},k_{-5})$ &(7.64,0.47)& (6.88,1.23)& (6.23,1.88)&(5.83,2.28)&(5.69,2.42)&  (5.61,2.50)& (5.57,2.54)& (5.48,2.63)&  (5.45,2.66) \\
 \noalign{\smallskip}
\cline{2-2}
 \noalign{\smallskip}
& $(k_{45},k_{-45})$ &(6.40,2.42)& (6.11,3.57)& (5.84,4.40)&(5.65,4.90)&(5.57,5.07)&  (5.53,5.16)& (5.50,5.22)& (5.45,5.33)&  (5.43,5.36) \\
        & $(k_f,k_r)^a$ &     &     &     &     &     &  (4.6,4.1)&     &     &  (4.1,4.1) \\
 \noalign{\smallskip}
\cline{2-11}
 \noalign{\smallskip}
D6 %$^{15}$N$^{14}$NH$^+$+ $^{15}$N$^{14}$N
            & $F_q$ & 0.9912& 0.9904 & 0.9901& 0.9899& 0.9898& 0.9898& 0.9898 & 0.9901& 0.9908\\
%$\rightleftharpoons ^{14}$N$^{15}$NH$^+$+ $^{14}$N$^{15}$N
            & $K_e^{(6)}$   &  5.21&  2.27&  1.50&  1.22&  1.14&  1.10&  1.08&  1.03&  1.02 \\  
& $(k_{6},k_{-6})$ &(6.80,1.31)& (5.63,2.48)& (4.87,3.24)&(4.45,3.66)&(4.31,3.80)&  (4.24,3.87)& (4.20,3.91)& (4.12,3.99)&  (4.09,4.02) \\
 \noalign{\smallskip}
\hline
%--------------------------------------------------------------
\end{tabular}
\tablefoot{
\tablefoottext{a}{\cite{adams81}}
}
\end{table*}
%-----------------------------------------------------------------------

%#######################

The equilibrium constants $K_e$ 
and rate coefficients
for the isotopic variants of N$_2$H$^+$ reacting with N$_2$ 
are shown in Table \ref{table_rate_n2h}.
There we assumed the total rate coefficient $k_{\mathrm{T}}$ 
given by the Langevin collision rate (in SI units)
\begin{eqnarray}
k_{\mathrm{L}} & = & e \sqrt{{\pi \, \alpha(\mathrm{N}_2)}/{\mu_R \varepsilon_0}}
,
% 2 \pi \mathrm{e} \sqrt{\alpha(A)/\mu_r}
\label{langevin}
\end{eqnarray}
\noindent
where $e$ is the elementary charge, $\mu_R$ the reduced mass for the collision,
and $\alpha(\mathrm{N}_2)$ the polarizability of N$_2$  
[$\alpha(\mathrm{N}_2)=1.710$ {\AA}$^3$ \citep{olney97}], giving thus
$k_{\mathrm{T}}=k_{\mathrm{L}} = 8.11 \times 10^{-10}$ cm$^3$s$^{-1}$.
The rate coefficients $k_f$ and $k_r$ are determined from $k_{\mathrm{T}}$ and $K_e$
with the help of Eqs. (\ref{kf_langer}) and (\ref{kr_langer}), respectively.
Spectroscopic parameters of \cite{trickl95} and \cite{bendtsen01} 
were used for the $X {^1\Sigma}^+_g$ states 
of $^{14}$N$_2$, $^{14}$N$^{15}$N, and $^{15}$N$_2$.

The nuclear spin degeneracy 
affects the equilibrium constants for the reactions
involving either $^{14}$N$_2$ or $^{15}$N$_2$. 
At higher temperatures, $K_e$ in Table \ref{table_rate_n2h} approaches 
1/2 for reactions D2 and D3
having ${^{14}\mathrm{N}}_2$ as a product,
and 2 for reactions D4 and D5
having ${^{15}\mathrm{N}}_2$ as a reactant. 
For reaction D1,
\begin{eqnarray}
{^{14}\mathrm{N}}_2{\mathrm{H}}^+
+ {^{15}\mathrm{N}}_2
\mathrel{\mathop{\rightleftharpoons}^{k_1}_{k_{-1}}} 
{^{15}\mathrm{N}}_2{\mathrm{H}}^+
+ {^{14}\mathrm{N}}_2,
\label{reaction_n2h_1}
\end{eqnarray}
\noindent
the effects from nuclear spin statistics cancel out
and $K_e^{(1)} \rightarrow 1$ as the temperature increases.
From Eqs. (\ref{n14}) and (\ref{n15}), 
the ortho-to-para ratio is
given by $R_{14}=6 Q_{\mathrm{evenJ}}/3 Q_{\mathrm{oddJ}}$ for $^{14}$N$_2$ 
and by $R_{15}=3 Q_{\mathrm{oddJ}}/ Q_{\mathrm{evenJ}}$ for $^{15}$N$_2$.
We may note that $R_{14}$ assumes a value of 2.41 (2.01)
and $R_{15}$ a value of 2.60 (2.99) at 5 K (10 K).
At high temperature equilibrium, we have $R_{14}=2$ and $R_{15}=3$.

\cite{adams81} employed normal nitrogen (ratio 2:1 of ortho vs para $^{14}$N$_2$)
in the SIFT experimental study of N$_2$H$^+$+N$_2$.
To measure the forward reaction and backward reaction rate coefficients
at a given temperature, they interchanged
the ion-source gas and reactant gas.
Using mass-selected samples, 
these authors, however, were unable to distinguish between 
the isotopomers $^{14}$N$^{15}$NH$^+$ and $^{15}$N$^{14}$NH$^+$,
such that their results provide the overall yield of these cations
(no information on the relative yields).
This applies to the competing reactions D2 and D3
on one side
and the competing reactions D4 and D5 on the other side.
$^{14}$N$^{15}$NH$^+$ and $^{15}$N$^{14}$NH$^+$
are expected to be differently fractionated
(see Table \ref{table_kelvin_n2h}).

To simulate the experimental conditions of \cite{adams81},
we introduced 
the overall forward $k_{23}$  and overall reverse $k_{-23}$ rate coefficients
for reactions D2 and D3, 
\begin{eqnarray}
k_{23} & = & k_2 + k_3,
\label{rate_23_f}\\
k_{-23} & = & 
\left[ k_{-2} {K_e^{(6)}} + k_{-3} \right] 
\frac{1}{1+K_e^{(6)}}
,
\label{rate_23_r}
\end{eqnarray}
\noindent
and the overall forward $k_{45}$ and overall reverse $k_{-45}$ rate coefficients 
for reactions D4 and D5,
\begin{eqnarray}
k_{-45} & = & k_{-4} + k_{-5},
\label{rate_45_r} \\
k_{45} & = & \left[ k_4 K_e^{(6)} + k_5 \right] \frac{1}{1 + K_e^{(6)}}
.
\label{rate_45_f} 
\end{eqnarray}
\noindent
Here we explicitly assumed
an equilibrium distribution between 
$^{14}$N$^{15}$NH$^+$ and $^{15}$N$^{14}$NH$^+$.
The term $1+K_e^{(6)}$ is the state-distribution normalization factor.

%-----------------------------------------------------------------------
\begin{figure}
 \resizebox{\hsize}{!}{\includegraphics{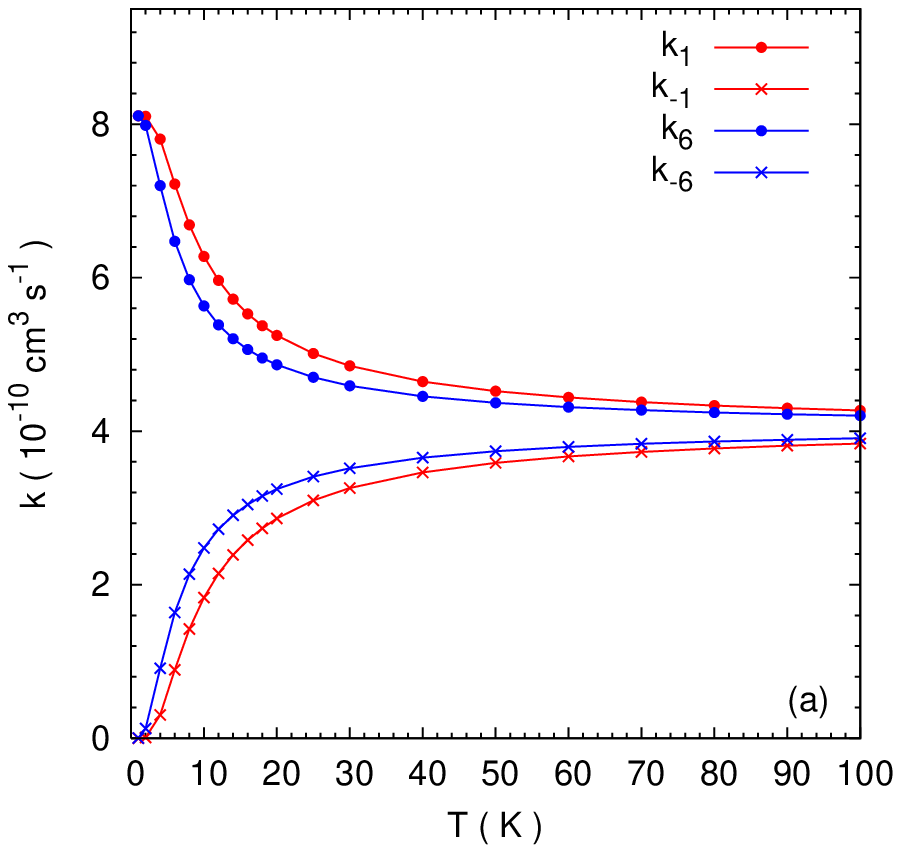}}
 \resizebox{\hsize}{!}{\includegraphics{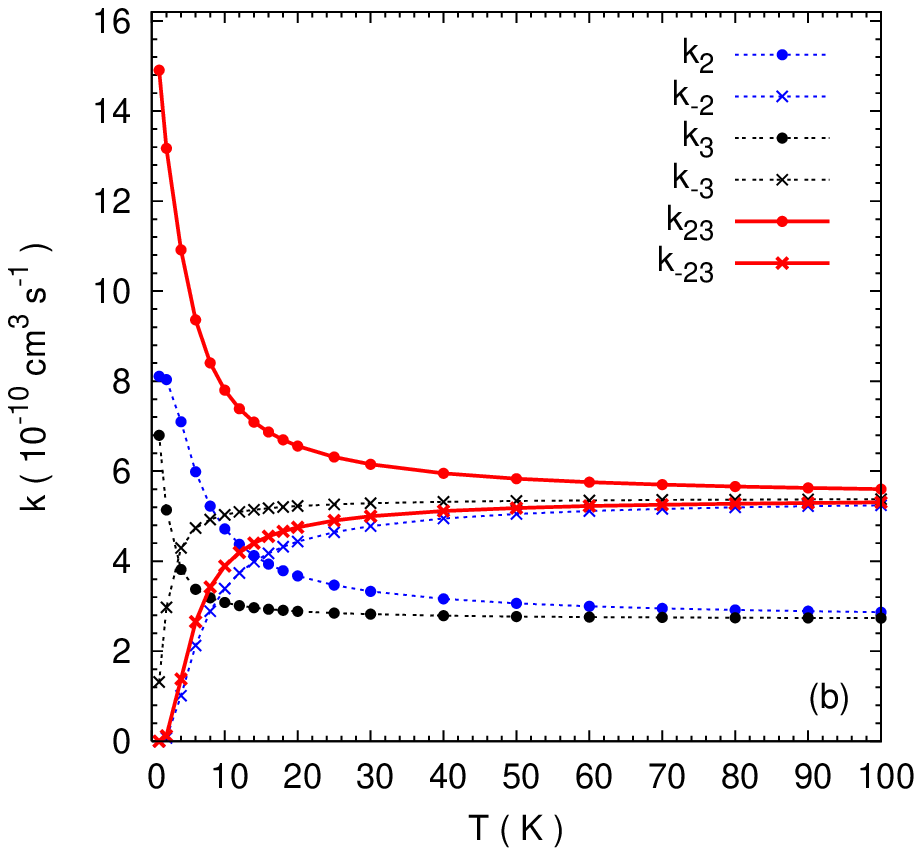}}
 \resizebox{\hsize}{!}{\includegraphics{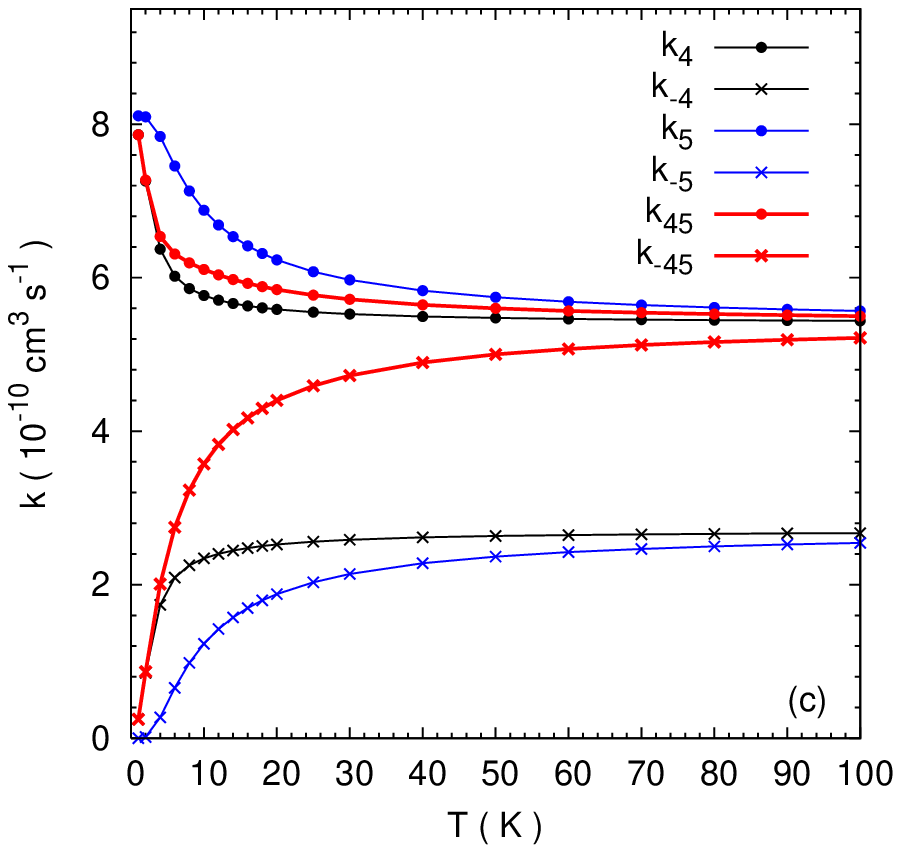}}
% \resizebox{\hsize}{!}{\includegraphics{rate16.eps}}
% \resizebox{\hsize}{!}{\includegraphics{rate23.eps}}
% \resizebox{\hsize}{!}{\includegraphics{rate45.eps}}
\caption{
Temperature dependence of the rate coefficients 
for $\mathrm{N}_2\mathrm{H}^++\mathrm{N}_2$
for reactions D1 and D6 in (a), reactions D2 and D3 in (b), 
and reactions D4 and D5 in (c).}
\label{fig.rate}
\end{figure}
%-----------------------------------------------------------------------

%-----------------------------------------------------------------------
\begin{figure}
%\includegraphics[width=0.50\textwidth]{fig2a.eps}
%\includegraphics[width=0.50\textwidth]{fig2b.eps}
%\vskip 0.25cm
%\includegraphics[width=0.50\textwidth]{fig2c.eps}
%\vskip 0.25cm
 \resizebox{\hsize}{!}{\includegraphics{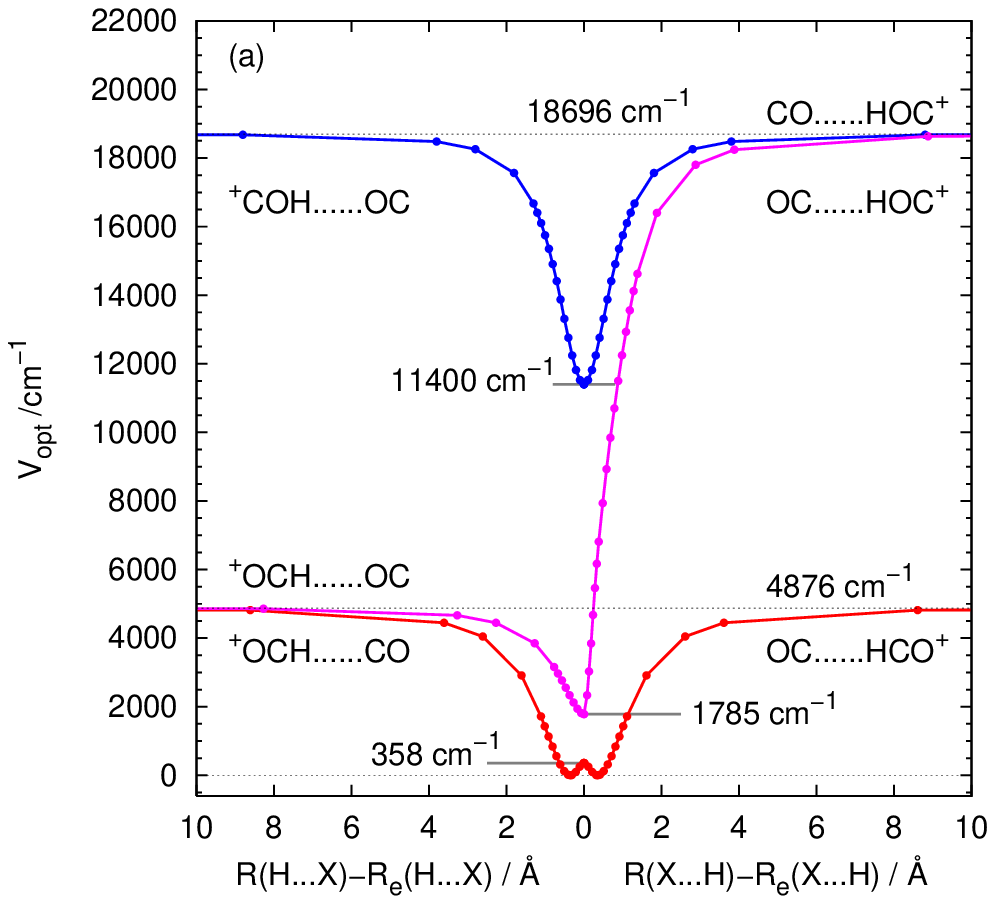}}
\vskip 0.25cm
 \resizebox{\hsize}{!}{\includegraphics{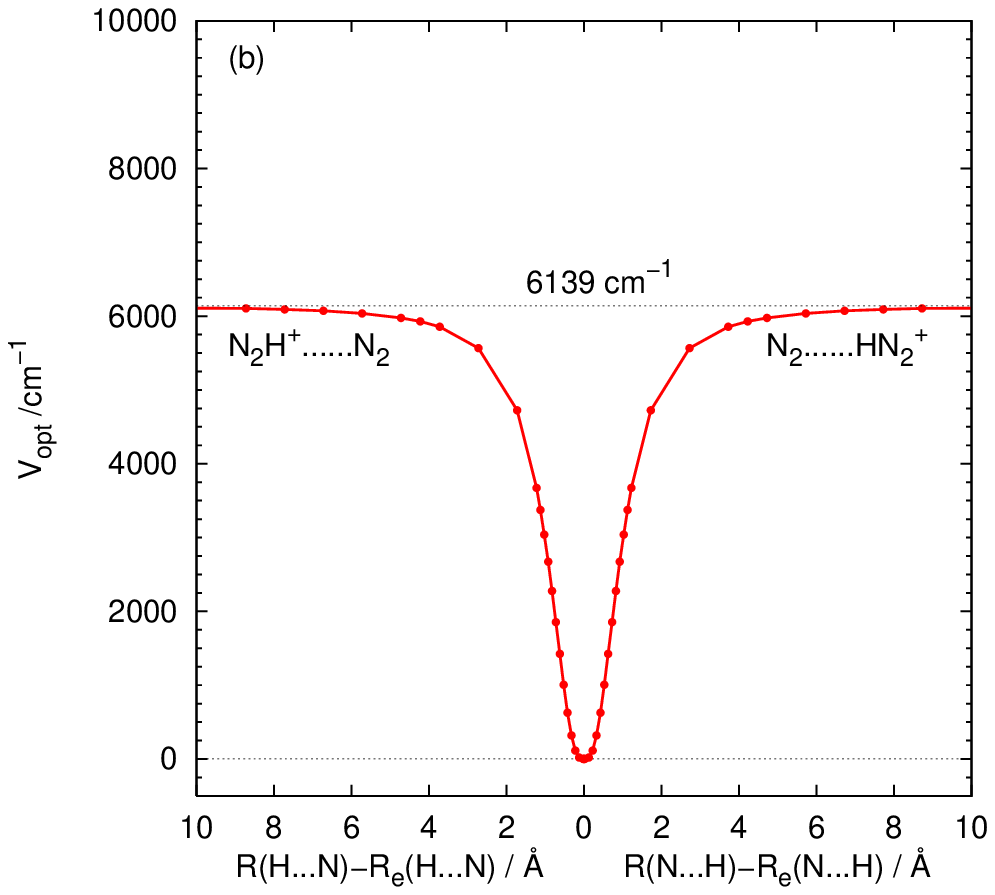}}
\vskip 0.25cm
 \resizebox{\hsize}{!}{\includegraphics{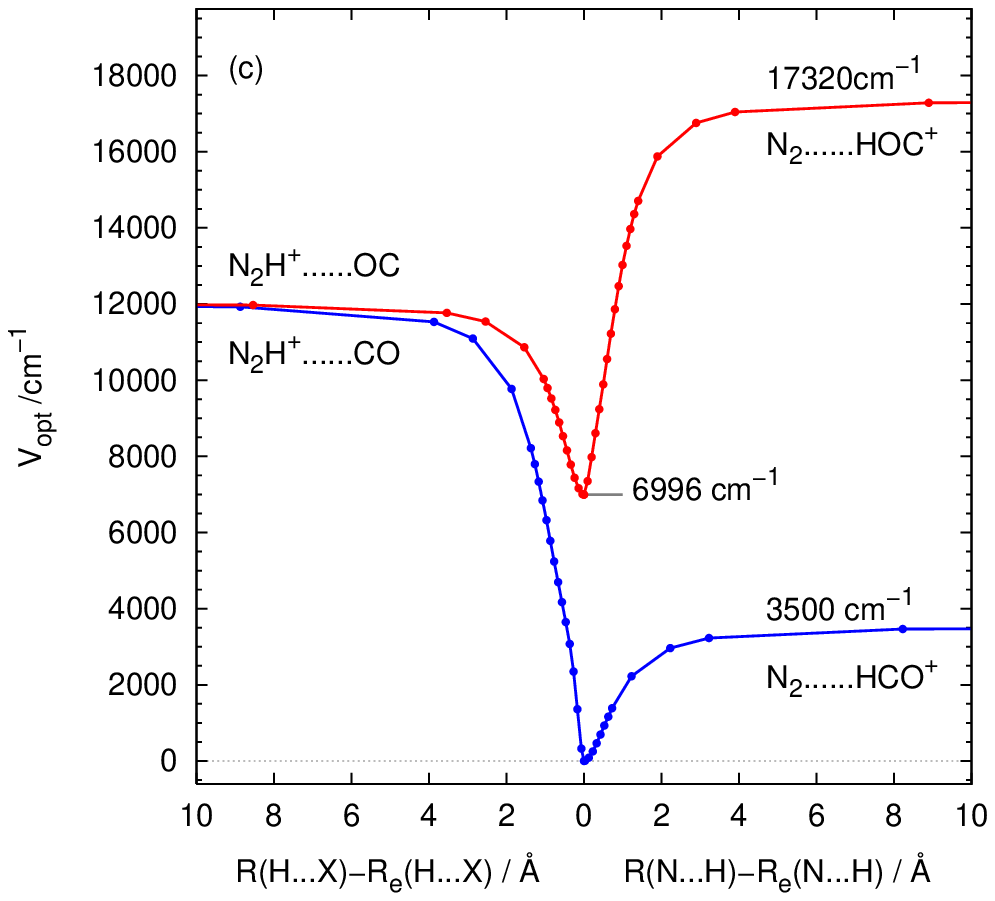}}
\vskip 0.25cm
% \resizebox{\hsize}{!}{\includegraphics{ochco-oba.eps}}
%\vskip 0.25cm
% \resizebox{\hsize}{!}{\includegraphics{n2hn2-oba.eps}}
%\vskip 0.25cm
% \resizebox{\hsize}{!}{\includegraphics{mix-oba.eps}}
%\vskip 0.25cm
\caption{
Minimum energy paths for the linear approach of CO to HCO$^+$/HOC$^+$
in (a), for the linear approach of N$_2$ to HN$_2^+$ in (b) and
for the formation of the mixed linear cluster ions 
N$_2\cdots$HCO$^+$ and N$_2$H$^+ \cdots$OC in (c).
The coordinate displayed on the \emph{x}-axis is shown with
the dotted line in the chemical formulas and X=O,C.
}
\label{fig.mep}
\end{figure}
%-----------------------------------------------------------------------

The variation of the rate coefficients with the temperature 
is displayed in Fig. \ref{fig.rate}.
The common feature seen there is that
the forward reaction becomes faster and
the backward reaction slower 
with decreaseing temperature.
%%as the temperature drops.
We also see that $k_f$ and $k_r$ exhibit 
a very weak temperature dependence for $T>50$ K.
For reactions D1 and D6, 
$k_f$ and $k_r$ approach the same value ($k_{\mathrm{L}}/2$ in our model)
at higher temperatures in Fig. \ref{fig.rate}(a) and 
Table \ref{table_rate_n2h}.
The high temperature limits of $k_i$ and $k_{-i}$ for $i=2-5$
are, however, different because of the nuclear spin restrictions,
as clearly seen in Figs. \ref{fig.rate}(b) and \ref{fig.rate}(c).
The reverse rate coefficients $k_{-2}$ and $k_{-3}$ in Fig. \ref{fig.rate}(b) 
become even higher than $k_{2}$ and $k_{3}$ for $T>14.7$ K and $T>3.4$ K,
respectively, inverting thus the reaction direction.
\cite{herbst03} also found $k_{-3} > k_3$ at $T=10$ K for 
reaction D3 assuming a different rate-coefficient model.
 
For the overall state-averaged rate coefficients
in Fig. \ref{fig.rate} and Table \ref{table_rate_n2h},
we have $k_{23} > k_{-23}$ and $k_{45}>k_{-45}$
for all temperatures shown. This is in accordance with
the SIFT experiment of \cite{adams81}. 
The rate coefficients $k_{\pm23}$ and $k_{\pm 45}$
appear 30\% higher than
the experimental finding, reported with
an error of $\pm$25\% at 80\,K.
Note, however, that
the ratios $k_{23}/k_{-23}$ and $k_{45}/k_{-45}$ agree within 6\%  
with the corresponding experimental values. 
Due to the nuclear spin angular momentum selection rules,
the high-temperature limits (for $K_e^{(6)} \rightarrow 1$)
of $k_{\pm 23}$ and $k_{\pm 45}$ 
are different 
from the high-temperature limits %of $k_{\mathrm{L}}/2$ 
of $k_{\pm 1}$ and $k_{\pm 6}$.

{\boldtext
From Eqs. (\ref{rate_23_f})-(\ref{rate_45_f}) and the relationship 
of Eq. (\ref{keq_relations}), we easily obtain
\begin{eqnarray}
\frac{k_{23}}{k_{-23}} & = & {K_e^{(2)}}+{K_e^{(3)}}
\end{eqnarray}
\noindent
and 
\begin{eqnarray}
\frac{k_{45}}{k_{-45}} & = & \frac{{K_e^{(4)}} {K_e^{(5)}}}{{K_e^{(4)}}+{K_e^{(5)}}}
.
\end{eqnarray}

\noindent
Following the procedure of \cite{adams81}, we may model the temperature dependence
of the latter ratios
as $e^{\Delta E_{ij}/k_{\mathrm{B}}T}$ [compare with Eq. (\ref{ratio_sa})].
Using our results from Table \ref{table_rate_n2h} 
for the overall forward and overall reverse rate coefficients  
calculated at the temperatures of the SIFT experimental study,
$T=80$ K and $T=292$ K, we derive 
$\Delta E_{23}, \Delta E_{45}=6.5$ K
for both reaction pairs. 

\cite{adams81} estimated the zero-point energy difference 
of $9 \pm 3$ K for reactions D2 and D4 
(see Table \ref{table_kelvin_n2h}). 
In accordance with the analysis presented here, we see, however, 
that the results of \cite{adams81} 
should be attributed to the reaction pairs 
$\{$D2,D3$\}$ and $\{$D4,D5$\}$.
This also explains a large discrepancy seen in Table \ref{table_kelvin_n2h}
between the theoretical estimates and experimental finding for reaction D4.
}

In recent studies of \cite{bizzocchi10,bizzocchi13},
$^{14}$N$^{15}$NH$^+$ and $^{15}$N$^{14}$NH$^+$ were both detected 
in a prototypical starless core L1544 of low central temperature %(T $<$ 15 K),
and an abundance ratio
$R^{14,15}_{15,14}=[{^{14}\mathrm{N}}{^{15}\mathrm{N}}\mathrm{H}^+]/[{^{15}\mathrm{N}}{^{14}\mathrm{N}}\mathrm{H}^+]$ 
of $1.1 \pm 0.3$ was derived.
%This abundance ratio nicely correlates with
%the values of $K_e^{(6)}$ for $T>20$ K in Table \ref{table_rate_n2h}.
%Note that the model of \cite{rodgers04}
%has led to an isotopologue abundance ratio of 1.8--2.3.
{\boldtext
%%$[{^{14}\mathrm{N}}{^{15}\mathrm{N}}\mathrm{H}^+]/[{^{15}\mathrm{N}}{^{14}\mathrm{N}}\mathrm{H}^+]$ 
Note that the ratio $R^{14,15}_{15,14}$ correlates with $K_e^{(6)}$ describing reaction D6 of
Eq. (\ref{reaction_n2h_6}).
As seen in Table \ref{table_rate_n2h}, 
we obtain $K_e^{(6)}$ of 1.22-1.02 for $T=40-292$ K; 
%we additionally calculated $R^{14,15}_{15,14}=1.31$ for T=30 K.
the additional calculation at T=30 K gave  $R^{14,15}_{15,14}=1.31$.
Also note that the earlier model of \cite{rodgers04}
has led to the ratio $R^{14,15}_{15,14}$ of 1.8--2.3, which correlates with
our $K_e^{(6)}$ value of 2.27 (1.72) at $T=10$\,K (15\,K).
}

%-------------------------------------------------------
\subsection{Ionic complexes}
\label{discussion:pes}

Ion-molecule reactions were additionally examined using 
electronic structure calculations, carried out
for the linear approach of the neutral CO and N$_2$ to the linear cations
HCO$^+$, HOC$^+$, and N$_2$H$^+$.
The corresponding minimum-energy paths (MEPs) are displayed in Fig. \ref{fig.mep}.
The MEPs are obtained optimizing three intramolecular distances 
for various monomer separations. 
Our calculations were performed at the CCSD(T)/aug-cc-pVTZ level
of theory 
{\boldtext employing the standard MOLPRO and CFOUR optimization/threshold
parameters.
}

The lower MEP in Fig. \ref{fig.mep}(c) is related to the reaction
\begin{eqnarray}
{\mathrm{N}}_2{\mathrm{H}}^+
+ \mathrm{CO}
%\mathrel{\mathop{\rightleftharpoons}
\rightarrow
\mathrm{HCO}^+
+ \mathrm{N}_2,
&&
\label{reaction_n2h_destruction}
\end{eqnarray}
\noindent
which is considered to be the main destruction path for N$_2$H$^+$
when CO is present in the gas phase at standard abundances [CO]/[H$_2$] $\sim 10^{-4}$
\citep{snyder77,jorgensen04}.
For this reaction, 
\cite{herbst75} reported a rate coefficient of 8.79$\times$10$^{-10}$ cm$^3$ s$^{-1}$
at 297$\pm$2 K.
No reverse reaction was detected \citep{anicich93}.
For reactions involving HOC$^+$,
\cite{freeman87} measured a rate coefficient $k$ of 6.70$\times$10$^{-10}$ cm$^3$ s$^{-1}$
for the following reaction
\begin{eqnarray}
\mathrm{HOC}^+ + \mathrm{CO}
\rightarrow
{\mathrm{HCO}}^+ + \mathrm{CO}
,
\label{reaction_hoc.2}
\end{eqnarray}
\noindent
whereas \cite{wagner-redeker85} reported $k$ as
6.70$\times$10$^{-10}$ cm$^3$ s$^{-1}$ for 
\begin{eqnarray}
\mathrm{HOC}^+ + \mathrm{N}_2
\rightarrow
{\mathrm{N}}_2{\mathrm{H}}^+ + \mathrm{CO}
.
\label{reaction_hoc}
\end{eqnarray}
\noindent
The Langevin collision rate 
is $k_{\mathrm{L}} = 8.67 \times 10^{-10}$ cm$^3$s$^{-1}$
for reactions (\ref{reaction_n2h_destruction}) and (\ref{reaction_hoc.2})
involving CO 
and $k_{\mathrm{L}} = 8.11 \times 10^{-10}$ cm$^3$s$^{-1}$
for reaction (\ref{reaction_hoc}) involving N$_2$.

%#######################
%\input{table_freq}
%--------------------------------------------------------------
\begin{table*}
\caption{\label{table_freq}
Properties of the ionic complexes from the CCSD(T)/aug-cc-pVTZ 
calculations.$^a$ }
\centering
\vspace{0.025cm}
%\tiny 
\begin{tabular}{lrrrrrrr}
\hline\hline
 \noalign{\smallskip}
           Quantity
         & {OCH$^+\cdots$CO}  
         & {OCHCO$^+$(TS)}  
         & {OCH$^+\cdots$OC} 
         & {COHOC$^+$} 
         & {N$_2 \cdots$HCO$^+$} 
         & {N$_2$H$^+\cdots$OC} & {N$_2$HN$_2^+$}   \\  
 \noalign{\smallskip}
\hline
 \noalign{\smallskip}
% \small
%%%%    & OCHCO$^+$ & TS         & CO$\cdots$HCO$^+$     & COHOC$^+$    & N$_2 \cdots$HCO$^+$ & N$_2$HOC$^+$ & N$_2$HN$_2^+$ \\
$r_1$ (\AA)& 1.116         & 1.121         & 1.114         &  1.153       &   1.103              &   1.099              & 1.101  \\
$r_2$ (\AA)& 1.175         & 1.387         & 1.118         &  1.194       &   1.774              &   1.104              & 1.276  \\
$r_3$ (\AA)& 1.741         & 1.387         & 1.729         &  1.194       &   1.131              &   1.460              & 1.276  \\
$r_4$ (\AA)& 1.125         & 1.121         & 1.146         &  1.153       &   1.114              &   1.149              & 1.101  \\
$B_e$ (cm$^{-1}$)& 0.0639          & 0.0681          &0.0706          & 0.0948          & 0.0670            &  0.0844               & 0.0827  \\
%%$E$/$E_h$& -226.581854   & -226.580220   & -226.573721   & -226.529906  & -222.794216          & -222.762345       & -218.986195   \\
%%$E$/$E_h$& -226.58185405 & -226.58022038 & -226.57372131 & -226.52990582& -222.79421633        & -222.76234525     & -218.98619506 \\
 \noalign{\smallskip}
$\omega_1$ (cm$^{-1}$)&  2\,464 (2\,397)  & 2\,306 (2\,305)   &  2\,887 (2\,520)  & 2\,081 (2\,081)  &      2\,745 (2\,480)      &   2\,555 (2\,466)  & 2\,402 (2\,402) \\
$\omega_2$ (cm$^{-1}$)&  2\,237 (2\,237)  & 2\,273 (2\,272)   &  2\,136 (2\,076)  & 2\,015 (2\,014)  &      2\,351 (2\,351)      &   2\,079 (2\,059)  & 2\,365  (2\,365) \\
$\omega_3$ (cm$^{-1}$)&  1\,730 (1\,291)  & 1\,290  (950)   &  2\,070 (1\,758)  & 1\,034  (745)  &      2\,077 (1\,658)      &   1\,945 (1\,474)    & 1\,235 (902) \\
$\omega_4$ (cm$^{-1}$)&  1\,145 (859)   & 1\,290  (950)   &   984  (760)  & 1\,034  (745)  &      1\,040 (794)       &   1\,039 (767)     & 1\,235 (902) \\
$\omega_5$ (cm$^{-1}$)&  1\,145 (859)   & 840i  (600i)  &   984   (760) &  988    (706)&      1\,040 (794)       &   1\,039 (767)     & 438   (438) \\
$\omega_6$ (cm$^{-1}$)&   271 (267)   & 395   (395)   &   193  (191)  &  467   (467) &       229 (221)      &    271 (267)         & 265  (265) \\
$\omega_7$ (cm$^{-1}$)&   271 (267)   & 295    (295)  &   189  (178)  &  134   (134) &       229 (221)      &    214 (209)         & 265  (265) \\
$\omega_8$ (cm$^{-1}$)&   199 (196)   & 295   (295)   &   189  (178)  &  134   (134) &       195 (193)      &    214  (209)        & 159 (112) \\
$\omega_9$ (cm$^{-1}$)&   131 (126)   & 148   (143)   &    91   (90)  &   89   (88)  &       117  (113)     &    103  (102)        & 144 (140) \\
$\omega_{10}$ (cm$^{-1}$)&131 (126)  & 148    (143)  &    91    (90) &   89    (88) &        117 (113)     &     103 (102)        & 144  (140) \\
$E_0$ (cm$^{-1}$)&   4\,862 (4\,313)  & 4\,220 (3\,875)  &   4\,908 (4\,300) & 4\,033 (3\,600)  &     5\,068 (4\,468)      &    4\,780 (4\,211)       & 4\,327 (3\,967) \\
 \noalign{\smallskip}
%$\ln Z$  & 4.689  (4.689)     & 3.932 (3.932) & 4.589  (4.590)& 3.601 (3.601)&    4.642 (4.642)     &    4.410 (4.410)     & 3.738 (3.738) \\
% \noalign{\smallskip}
$\mu_e$ ($ea_0$)&   1.157       &    0          &  1.364        &     0        &     -1.389           &    0.808             &  0 \\
$\Theta_{zz}$ ($ea_0^2$)& 8.100      &  7.302        &   7.759       &   8.363      &      8.284           &    8.252             &  8.108 \\
 \noalign{\smallskip}
\hline
%--------------------------------------------------------------
\end{tabular}
\tablefoot{
\tablefoottext{a}{Wavenumbers for the deuterated species are given in parentheses.} 
}
\end{table*}
%-----------------------------------------------------------------------

%#######################

%#######################
%\input{table_freq-monomers}
%--------------------------------------------------------------
\begin{table*}
\caption{\label{table_freq-monomers}
Properties of the monomers CO, N$_2$, HCO$^+$, HOC$^+$, and N$_2$H$^+$ 
from the CCSD(T)/aug-cc-pVTZ calculations.$^{a,b}$ }
\centering
\vspace{0.025cm}
%\tiny 
\begin{tabular}{lccccc}   
\hline\hline
 \noalign{\smallskip}
           Quantity
         & {CO}
         & {HCO$^+$} & {HOC$^+$} 
         & {N$_2$}
         & {N$_2$H$^+$} 
  \\  
 \noalign{\smallskip}
\hline
 \noalign{\smallskip}
%%        &   CO          & HCO           &  HOC          &    N2         &   N2H
$r_1$ (\AA) & 1.136         &  1.094        &  0.992        &   1.104       &  1.099              \\
$r_2$ (\AA) &               &  1.113        &  1.162        &               &  1.035              \\
%$B_e$/MHz & 57118         & 44328 (35809) & 44247 (37620) &  59222        & 46352 (38350)   \\
%$E$/$E_h$ & -113.162194   & -113.397431   & -113.334466   & -109.380845   & -109.577381      \\
%%%%$E$/$E_h$ & -113.16219365 & -113.39743063 & -113.33446568 & -109.38084518 & -109.57738138    \\
 \noalign{\smallskip}
$\omega_1$ (cm$^{-1}$)&    2\,144     &  3\,211 (2\,634)  &  3\,471 (2\,592)  &   2\,339        &  3\,395 (2\,711)   \\
$\omega_2$ (cm$^{-1}$)&               &  2\,192 (1\,923)  &  1\,931 (1\,861)  &               &  2\,276 (2\,052)   \\
$\omega_{3,4}$ (cm$^{-1}$)&               &   844 (676)   &    61  (48)   &               &   686  (543)    \\
%%$\omega_4$ (cm$^{-1}$)&               &   844 (676)   &    61  (48)   &               &   686  (543)    \\
$E_0$ (cm$^{-1}$)     & 1\,072        &  3\,546 (2\,954)  & 2\,762 (2\,275)   &   1\,170        & 3\,522  (2\,924)   \\
 \noalign{\smallskip}
$\mu_e$ ($ea_0$) &  -0.040            & 1.537         & 1.083         &    0          & -1.328    \\
$\Theta_{zz}$ ($ea_0^2$) 
              & -1.466                & 4.235         & 4.198         &   -1.106      &  4.447    \\
 \noalign{\smallskip}
\hline
\end{tabular}
% \noalign{\smallskip}
%--------------------------------------------------------------
\tablefoot{
\tablefoottext{a}{Wavenumbers for the deuterated species are given in parentheses.}
\tablefoottext{b}{Vibrational modes $(\omega_3,\omega_4)$ are doubly degenerate.} 
}
\end{table*}
%-----------------------------------------------------------------------

%#######################

The common feature in Fig. \ref{fig.mep} is the formation of 
a linear proton-bound ionic complex,
which is 2000-7000 {\cm} more stable than the separated monomers.
The properties of the complexes are summarized 
in Table \ref{table_freq}, where we give
the geometric parameters $r_i$, the equilibrium rotational constants $B_e$,
the harmonic wavenumbers $\omega_i$ for the main and deuterated isotopologues, 
and the harmonic zero-point energies $E_0$.
%%are listed in Table \ref{table_freq} for the linear ionic complexes and
The corresponding results for the constituent monomers are listed 
in Table \ref{table_freq-monomers}.
Note that the monomer values $E_0$ in Table \ref{table_freq-monomers} are harmonic 
and therefore different from the anharmonic results of Table \ref{table_zpe}.
The coordinates $r_i (i=1-4)$ for A--B--H--C--D denote    
$r_1=r(\mathrm{A-B}), r_2=r(\mathrm{B-H}), r_3=r(\mathrm{H-C})$ 
and $r_4=r(\mathrm{C-D})$ 
in Table \ref{table_freq} and similar in Table \ref{table_freq-monomers}.
The dipole moments $\mu_z$ and the quadrupole moments $\Theta_{zz}$
in Table \ref{table_freq} and \ref{table_freq-monomers}
are given with respect to the inertial reference frame with the origin in the complex
centre of mass, where the position of the first atom A of A--B--H--C--D or A--B--C 
along the $z$ axis is chosen to be the most positive. 

The ionic complexes N$_2$HN$_2^+$ and COHOC$^+$ have linear 
centrosymmetric equilibrium structures. The complex OCH$^+$$\cdots$CO
is asymmetric with a barrier height to the centrosymmetric saddle point 
OCHCO$^+$(TS), seen at 358 {\cm} in Fig. \ref{fig.mep}(a).
In the mixed-cluster ions, the proton is bound either to CO, when
N$_2\cdots$HCO$^+$ is formed, or to N$_2$, when N$_2$H$^+$$\cdots$OC is formed.
Comparison of Tables \ref{table_freq} and \ref{table_freq-monomers}
%%indicates prominent changes of the geometric parameters (up to 0.01--0.02 \AA)
shows that the geometric parameters experience prominent changes (up to 0.01--0.02 \AA)
upon complexation.
% with the distances changing by approximately 0.01-0.02 {\AA} except $r($N-N).
In this fashion, the ionic (molecular) complexes differ from van der Waals complexes, 
in which the monomers preserve their geometric parameters to a great extent. 

%###########################
%\input{table_freq-complex}
%--------------------------------------------------------------
\begin{table*}
\caption{\label{table_freq-complex}
Isotopic variants of the ionic complex OCH$^+\cdots$CO.$^{a,b}$
}
\centering
\vspace{0.025cm}
\begin{tabular}{lcccccc}
\hline\hline
 \noalign{\smallskip}
          Quantity
         & {16-12-H+13-16}  
         & {18-12-H+13-18}  
         & {16-12-H+12-18}  
         & {16-13-H+13-18}  
         & {16-12-H+13-18}  
         & {18-12-H+13-16}  
\\
         & {reaction F1}
         & {reaction F2}
         & {reaction F3}
         & {reaction F4}
         & {reaction F5}
         & {reaction F6}
\\
 \noalign{\smallskip}
\hline
 \noalign{\smallskip}
$\omega_1$ (cm$^{-1}$)    %&  2464 
                  & 2\,464   & 2\,426   & 2\,464   & 2\,409   & 2\,464   & 2\,426 \\ 
$\omega_2$ (cm$^{-1}$)   %&  2237 
                  & 2\,187   & 2\,133   & 2\,185   & 2\,133   & 2\,133   & 2\,187 \\
$\omega_3$ (cm$^{-1}$)   %&  1730 
                  & 1\,730   & 1\,712   & 1\,731   & 1\,728   & 1\,730   & 1\,712 \\
$\omega_{4,5}$ (cm$^{-1}$)  %&  1145 
                   & 1\,145   & 1\,144   & 1\,147   & 1\,141   & 1\,149   & 1\,144 \\
%$\omega_5$ (cm$^{-1}$)    %&  1145 
%                   & 1\,145   & 1\,144   & 1\,147   & 1\,141   & 1\,149   & 1\,144 \\
$\omega_{6,7}$ (cm$^{-1}$)   %&   271 
                   &  267   &  264   &  271   &  261   &  266   &  265 \\
%$\omega_7$ (cm$^{-1}$)   %&   271 
%                   &  267   &  264   &  271   &  261   &  266   &  265 \\
$\omega_8$ (cm$^{-1}$)  %&   199 
                   &  198   &  191   &  197   &  193   &  194   &  195 \\
$\omega_{9,10}$ (cm$^{-1}$)   %&   131 
                   &  129   &  126   &  131   &  126   &  127   &  128 \\
%$\omega_{10}$ (cm$^{-1}$) %&   131 
%                   &  129   &  126   &  131   &  126   &  127   &  128 \\
 \noalign{\smallskip}
$B_e$ (MHz)         %& 1916  
                   & 1\,901   & 1\,728   & 1\,826   & 1\,799   & 1\,814   & 1\,812 \\
%$B_e$/\cm         & 0.6391& 0.6341 &0.5764  &0.6091  & 0.6001 & 0.6051 & 0.6044 \\
%$\ln Z_{10}$      & 12.845& 12.879 & 13.075 & 12.945 & 13.010 & 12.977 & 12.978 \\  
%$\ln Z_{100}$     & 20.904& 20.938 & 21.134 & 21.004 & 21.070 & 21.036 & 21.037 \\  
 \noalign{\smallskip}
$E_0$ (cm$^{-1}$)        %&  4862 
                   & 4\,830   & 4\,765   & 4\,836   & 4\,760   & 4\,798   & 4\,796 \\
$E_0^{r}$ (cm$^{-1}$)    %&  4618 
                   & 4\,594   & 4\,537   & 4\,592   & 4\,531   & 4\,568   & 4\,563  \\
$E_0^p$ (cm$^{-1}$)      %&       
                   & 4\,581   & 4\,524   & 4\,587   & 4\,526   & 4\,550   & 4\,555 \\
 \noalign{\smallskip}
$\Delta E^h/k_{\mathrm{B}}$ (K)     & 18.3   & 18.3   &  6.6   & 6.6    &  24.7  & 11.7 \\
%$\Delta E^h$ /\cm    & 13     & 13     &  5     & 5      &  17    & 8    \\
$D_0^r$ (cm$^{-1}$)        & 4\,640   & 4\,648   & 4\,632   & 4\,647   & 4\,645   & 4\,643 \\ 
$D_0^p$ (cm$^{-1}$)       & 4\,627   & 4\,635   & 4\,627   & 4\,642   & 4\,628   & 4\,635   \\ 
 \noalign{\smallskip}
\hline
%--------------------------------------------------------------
\end{tabular}
\tablefoot{
\tablefoottext{a}{Here, $a$-$b$-H+$c$-$d$ stands for
$^a$O$^b$CH$^+\cdots$$^c$C$^d$O.}
\tablefoottext{b}{Vibrational modes $(\omega_4,\omega_5)$, $(\omega_6,\omega_7)$,
and $(\omega_9,\omega_{10})$ are doubly degenerate.} 
}
\end{table*}
%----------------------------------------------------------------------

%###########################

The transformations in Fig. \ref{fig.mep} are all of the proton transfer type.
The neutral CO may approach H$^+$ of the triatomic cation 
either with C or O since both C and O possess lone electron pairs.
%% to attach the proton.  
The proton attachment from the C side leads to a more stable complex. 
As seen in Fig. \ref{fig.mep}(a),
the complex OCH$^+$$\cdots$OC is 1\,785 {\cm} above OCH$^+$$\cdots$CO
and 9\,615 {\cm} below COHOC$^+$. 
We also see that N$_2$$\cdots$HCO$^+$
is 6\,996 {\cm} more stable than N$_2$H$^+$$\cdots$OC. 
In all cases,
the energy separation between the HCO$^+$- and HOC$^+$-containing complexes
is smaller than the separation between free HCO$^+$ and HOC$^+$, seen to be
13\,820 {\cm} in Fig. \ref{fig.mep}.
The results of Fig. \ref{fig.mep} are consistent with the fact that
the proton tends to localize on the species with higher proton affinity.
The experimental proton affinity
is 594 kJ mol$^{-1}$ (49\,654 cm$^{-1}$) for CO on the C end 
and 427 kJ mol$^{-1}$ (35\,694 cm$^{-1}$) for CO on the O end 
\citep{freeman87}.
The experimental proton affinity of 
498 kJ mol$^{-1}$ (41\,629 cm$^{-1}$) 
was determined for N$_2$ \citep{ruscic91}.
% P.A.(N$_2$)=4.93$\pm$0.11 eV \citep{foner78}.

The harmonic wavenumbers for the ionic complexes occurring in the course of 
reactions F1--F6 are provided in Table \ref{table_freq-complex}.
In addition to the spectroscopic properties, we also give
the harmonic zero-point energies of the complexes $E_0$, 
the reactants $E_0^r$, and the products $E_0^p$, as well as
the dissociation energies including the harmonic zero-point energy correction 
in the direction of the reactants, $D_0^r=D_e+E_0^r-E_0$, 
and in the direction of the products, $D_0^p=D_e+E_0^p-E_0$, 
where $D_e$ is the classical dissociation energy. 
 
In Table \ref{table_freq-complex}, 
the vibrational mode $\omega_2$, which is predominantly
the diatom CO stretching vibration, is the most sensitive to
isotopic substitutions. 
Compared with $\omega$ of free CO,
$\omega_2$ exhibits a blue-shift of 93 {\cm} for the main isotopologue
(Table \ref{table_freq-monomers} vs. Table \ref{table_freq}).
The modes $\omega_1$ and $\omega_3$, highly sensitive to 
the H$\rightarrow$D substitution (Table \ref{table_freq}), 
can be considered as the H-C-O stretching modes.
The intermolecular stretching mode is $\omega_8$.
The zero-point-corrected dissociation energies in Table \ref{table_freq-complex}
are approximately 240 {\cm} lower than the electronic dissociation energy
of 4\,876 cm$^{-1}$ (Fig. \ref{fig.mep}). 
The harmonic $\Delta E^h/k_{\mathrm{B}}$ values 
in Table \ref{table_freq-complex} and anharmonic $\Delta E/k_{\mathrm{B}}$ values
in Table \ref{table_kelvin} agree within 0.5 K.

The proton-bound complexes OCH$^+\cdots$CO and
N$_2 \cdots$HCO$^+$ have large dipole moments $\mu_e$
of 2.94 D and 3.53 D (Table \ref{table_freq}).
%%of respectively $\mu_e =2.94$ D and $\mu_e=3.53$ D (Table \ref{table_freq}).
For OCH$^+\cdots$CO, the most intense infrared transitions are expected 
for $\omega_3$ (with harmonic intensity
$I^h_3$ of 2\,440 km mol$^{-1}$) and $\omega_1$ ($I^h_1=536$ km mol$^{-1}$),
whereas the intermolecular stretch $\omega_8$ has $I^h_8=232$ km mol$^{-1}$.
The fundamental (anharmonic) transitions
$(\nu_1,\nu_2,\nu_3,\nu_{4,5},\nu_{6,7},\nu_8,\nu_{9,10})$
are calculated to be
(2\,267, 2\,236, 1\,026, 1\,136, 346, 186, 208) for the main isotopologue
(in {\cm}). 
The most intense infrared active transitions for N$_2 \cdots$HCO$^+$ are
$\omega_1$ ($I^h_1=1\,034$ km mol$^{-1}$),
$\omega_3$ ($I^h_3=814$ km mol$^{-1}$),
and $\omega_8$ ($I^h_8=154$ km mol$^{-1}$).
For this complex, the fundamental vibrational 
$(\nu_1,\nu_2,\nu_3,\nu_{4,5},\nu_{6,7},\nu_8,\nu_{9,10})$
transitions are determined to be
(2\,357, 2\,321, 1\,876, 1\,045, 127, 186, 113)
(in {\cm}). 
%The infrared active transitions for N$_2$HN$_2^+$ are
%$\omega_2$ ($I_1=152$ km mol$^{-1}$),
%$\omega_8$ ($I_8=5516$ km mol$^{-1}$).
%
The anharmonic transitions are calculated
from the cubic and semi-diagonal quartic force field 
in a normal coordinate representation
by means of vibrational second-order perturbation theory,
as implemented in CFOUR \citep{CFOUR_brief}.

{\boldtext 
Regarding the CCSD(T)/aug-cc-pVTZ method used here, we may note that 
our value of 358 {\cm} in Fig. \ref{fig.mep}(a) 
for the barrier height of OC+HCO$^+$ 
agrees reasonably well
with previous theoretical results
of 382 {\cm} [the CCSD(T)/cc-pVQZ approach of \cite{botschwina01}]
and 398 {\cm} [the CCSD(T)/aug-cc-pVXZ approach of \cite{terrill10}
at the complete basis-set limit]. 
A classical dissociation energy 
was previously determined to be
4\,634 {\cm} for OCH$^+\cdots$CO 
and 5\,828 {\cm} for N$_2$HN$_2^+$ 
at the complete basis-set limit \citep{terrill10}.}
The use of larger basis sets would ultimately be needed for converging 
theoretical results to stable values.
{\boldtext Our primary goal here is 
the acquisition of first information relevant for the physical behaviour  
of the ionic complexes involving HCO$^+$ and N$_2$H$^+$. 
For these initial explorations of the potential energy surfaces,
the CCSD(T)/aug-cc-pVTZ approach is of satisfactory quality.
A more detailed analysis of various basis-set effects,
including the basis-set superposition error in systems with
significantly deformed monomers,
is being prepared 
and will be presented elsewhere.
}

%Regarding the CCSD(T)/aug-cc-pVTZ method used here, we may mention
%that the barrier height for OC+HCO$^+$ was previously found 
%to be 382 {\cm} in the CCSD(T)/cc-pVQZ approach \citep{botschwina01}
%and 398 {\cm} at the complete basis set limit \citep{terrill10}.
%{\boldtext 
%Our value of 358 {\cm} in Fig. \ref{fig.mep}(a) is in 
%A classical dissociation energy at the complete basis set limit 
%was determined to be
%4\,634 {\cm} for OCH$^+\cdots$CO 
%and 5\,828 {\cm} for N$_2$HN$_2^+$ \citep{terrill10};
%our respective values are
%4\,876 {\cm} [Fig. \ref{fig.mep}(a)]
%and 6\,139 {\cm} [Fig. \ref{fig.mep}(b)].
%}
%The use of larger basis sets would ultimately be needed for converging 
%theoretical results to stable values.
%However, our primary goal is 
%the acquisition of first information relevant for the physical behaviour  
%of the ionic complexes studied here. In connection with this, we may note that 
%the CCSD(T)/aug-cc-pVTZ approach is of satisfactory quality
%for initial explorations of the potential energy surfaces.
%A more detailed analysis is underway.

%------------------------------------------------------------------------------------
\section{Conclusion}
\label{sec:conclusion}

Ion-molecule reactions are common in interstellar space,
and investigating them helps to quantitatively 
understand the molecular universe \citep{watson76a}.
%%In the present work, 
We studied the isotope fractionation reactions 
of HCO$^+$/HOC$^+$ with CO and N$_2$H$^+$ with N$_2$, as well as
the linear proton-bound complexes formed in the course of
these reactions.
For OCH$^+$+CO, we pointed out inaccuracies of previous 
exothermicity values that are
commonly employed in chemical networks. 
The new exothermicities affect particularly prominently
the rate coefficients derived at temperatures 
of dark interstellar cloud environments,
which markedly changes the abundance ratios
of the $^{13}$C- and $^{18}$O-containing formyl isotopologues.

The linear proton-bound cluster ions are found to be 
strongly bound (2\,000--7\,000 \cm).
The ionic complexes OCH$^+\cdots$CO and OCH$^+\cdots$N$_2$
have sizeable dipole moments (2.9-3.5 D) and rotational constants of
approximately 2\,000 MHz.
If stabilized by means of collision and/or radiative processes, their high
rotational population may facilitate the detection of these ions at low temperatures.

\begin{acknowledgements}
MM is grateful to Geerd~H.~F. Diercksen for
sending her a copy of the MPI/PAE Astro 135 report.
Marius Lewerenz is acknowledged for helpful discussions.
Mila Lewerenz is thanked for helping with the literature search.
\end{acknowledgements}

%............................................................................
%\bibliographystyle{aa}
%\bibliography{mirjana,malw}

\newcommand{\noopsort}[1]{} \newcommand{\printfirst}[2]{#1}
  \newcommand{\singleletter}[1]{#1} \newcommand{\switchargs}[2]{#2#1}

%..............................................................................

%-------------------------------------------------------------------
%\input{table_zpe}
%\input{table_kelvin}
%\input{table_zpe_n2h}
%\input{table_kelvin_n2h}
%\input{table_rate}
%\input{table_evelyne_final}
%\input{table_rate_n2h}
%\input{table_freq}
%\input{table_freq-monomers}
%\input{table_freq-complex}
%-------------------------------------------------------------------

\end{document}